\documentclass[aps,twocolumn,superscriptaddress,groupedaddress]{revtex4}
\pdfoutput=1

\usepackage{graphicx}
\usepackage{dcolumn}
\usepackage{bm}

\usepackage{appendix}
\usepackage{tikzit}
\usepackage{tikz, circuitikz}
\tikzstyle{env}=[copoint,regular polygon rotate=0,minimum width=0.2cm, fill=black]

\tikzstyle{probs}=[shape=semicircle,fill=white,draw=black,shape border rotate=180,minimum width=1.2cm]

\tikzstyle{nudge}=[yshift=0.6mm]

\tikzstyle{every picture}=[baseline=-0.25em,scale=0.5]
\tikzstyle{dotpic}=[] 
\tikzstyle{diredges}=[every to/.style={diredge}]
\tikzstyle{math matrix}=[matrix of math nodes,left delimiter=(,right delimiter=),inner sep=2pt,column sep=1em,row sep=0.5em,nodes={inner sep=0pt},text height=1.5ex, text depth=0.25ex]

\tikzstyle{inline text}=[text height=1.5ex, text depth=0.25ex,yshift=0.5mm]
\tikzstyle{label}=[font=\footnotesize,text height=1.5ex, text depth=0.25ex]
\tikzstyle{left label}=[label,anchor=east,xshift=2mm]
\tikzstyle{right label}=[label,anchor=west,xshift=-2mm]

\tikzstyle{braceedge}=[decorate,decoration={brace,amplitude=2mm,raise=-1mm}]
\tikzstyle{small braceedge}=[decorate,decoration={brace,amplitude=1mm,raise=-1mm}]

\tikzstyle{doubled}=[line width=1.6pt] 
\tikzstyle{boldedge}=[doubled,shorten <=-0.17mm,shorten >=-0.17mm]
\tikzstyle{boldedgegray}=[doubled,gray,shorten <=-0.17mm,shorten >=-0.17mm]
\tikzstyle{singleedgegray}=[gray]

\tikzstyle{semidoubled}=[line width=1.4pt] 
\tikzstyle{semiboldedgegray}=[semidoubled,gray,shorten <=-0.17mm,shorten >=-0.17mm]

\tikzstyle{boxedge}=[semiboldedgegray]

\tikzstyle{boldedgedashed}=[very thick,dashed,shorten <=-0.17mm,shorten >=-0.17mm]
\tikzstyle{vboldedgedashed}=[doubled,dashed,shorten <=-0.17mm,shorten >=-0.17mm]
\tikzstyle{left hook arrow}=[left hook-latex]
\tikzstyle{right hook arrow}=[right hook-latex]
\tikzstyle{sembracket}=[line width=0.5pt,shorten <=-0.07mm,shorten >=-0.07mm]

\tikzstyle{causal edge}=[->,thick,gray]
\tikzstyle{causal nondir}=[thick,gray]
\tikzstyle{timeline}=[thick,gray, dashed]

\tikzstyle{cedge}=[<->,thick,gray!70!white]

\tikzstyle{empty diagram}=[draw=gray!40!white,dashed,shape=rectangle,minimum width=1cm,minimum height=1cm]
\tikzstyle{empty diagram small}=[draw=gray!50!white,dashed,shape=rectangle,minimum width=0.6cm,minimum height=0.5cm]

\tikzstyle{dot}=[inner sep=0mm,minimum width=2mm,minimum height=2mm,draw,shape=circle]  
\tikzstyle{Wsquare}=[white dot, shape=regular polygon, rounded corners=0.8 mm, minimum size=3.3 mm, regular polygon sides=3, outer sep=-0.2mm]
\tikzstyle{Wsquareadj}=[white dot, shape=regular polygon, rounded corners=0.8 mm, minimum size=3.3 mm, regular polygon sides=3, outer sep=-0.2mm, regular polygon rotate=180]
\tikzstyle{ddot}=[inner sep=0mm, doubled, minimum width=2.5mm,minimum height=2.5mm,draw,shape=circle]

\tikzstyle{black dot}=[dot,fill=black]
\tikzstyle{white dot}=[dot,fill=white,,text depth=-0.2mm]
\tikzstyle{white Wsquare}=[Wsquare,fill=white,,text depth=-0.2mm]
\tikzstyle{white Wsquareadj}=[Wsquareadj,fill=white,,text depth=-0.2mm]
\tikzstyle{green dot}=[white dot] 
\tikzstyle{gray dot}=[dot,fill=gray!40!white,,text depth=-0.2mm]
\tikzstyle{red dot}=[gray dot] 

\tikzstyle{black ddot}=[ddot,fill=black]
\tikzstyle{white ddot}=[ddot,fill=white]
\tikzstyle{gray ddot}=[ddot,fill=gray!40!white]

\tikzstyle{gray edge}=[gray!60!white]

\tikzstyle{small dot}=[inner sep=0.5mm,minimum width=0pt,minimum height=0pt,draw,shape=circle]

\tikzstyle{small black dot}=[small dot,fill=black]
\tikzstyle{small white dot}=[small dot,fill=white]
\tikzstyle{small gray dot}=[small dot,fill=gray!40!white]

\tikzstyle{very small dot}=[inner sep=0.3mm,minimum width=0pt,minimum height=0pt,draw,shape=circle]

\tikzstyle{very small black dot}=[very small dot,fill=black]
\tikzstyle{very small white dot}=[small dot,fill=white]
\tikzstyle{very small gray dot}=[small dot,fill=gray!40!white]

\tikzstyle{causal dot}=[inner sep=0.4mm,minimum width=0pt,minimum height=0pt,draw=white,shape=circle,fill=gray!40!white]

\tikzstyle{phase dimensions}=[minimum size=5mm,font=\footnotesize,rectangle,rounded corners=2.5mm,inner sep=0.2mm,outer sep=-2mm]
\tikzstyle{dphase dimensions}=[minimum size=5mm,font=\footnotesize,rectangle,rounded corners=2.5mm,inner sep=0.2mm,outer sep=-2mm]

\tikzstyle{white phase dot}=[dot,fill=white,phase dimensions]
\tikzstyle{white phase ddot}=[ddot,fill=white,dphase dimensions]

\tikzstyle{white rect ddot}=[draw=black,fill=white,doubled,minimum size=5mm,font=\footnotesize,rectangle,rounded corners=2.5mm,inner sep=0.2mm]
\tikzstyle{gray rect ddot}=[draw=black,fill=gray!40!white,doubled,minimum size=6mm,font=\footnotesize,rectangle,rounded corners=3mm]

\tikzstyle{gray phase dot}=[dot,fill=gray!40!white,phase dimensions]
\tikzstyle{gray phase ddot}=[ddot,fill=gray!40!white,dphase dimensions]
\tikzstyle{grey phase dot}=[gray phase dot]
\tikzstyle{grey phase ddot}=[gray phase ddot]

\tikzstyle{small phase dimensions}=[minimum size=4mm,font=\tiny,rectangle,rounded corners=2mm,inner sep=0.2mm,outer sep=-2mm]
\tikzstyle{small dphase dimensions}=[minimum size=4mm,font=\tiny,rectangle,rounded corners=2mm,inner sep=0.2mm,outer sep=-2mm]

\tikzstyle{small gray phase dot}=[dot,fill=gray!40!white,small phase dimensions]
\tikzstyle{small gray phase ddot}=[ddot,fill=gray!40!white,small dphase dimensions]

\tikzstyle{small map}=[draw,shape=rectangle,minimum height=4mm,minimum width=4mm,fill=white]

\tikzstyle{cnot}=[fill=white,shape=circle,inner sep=-1.4pt]

\tikzstyle{asym hadamard}=[fill=white,draw,shape=NEbox,inner sep=0.6mm,font=\footnotesize,minimum height=4mm]
\tikzstyle{asym hadamard conj}=[fill=white,draw,shape=NWbox,inner sep=0.6mm,font=\footnotesize,minimum height=4mm]
\tikzstyle{asym hadamard dag}=[fill=white,draw,shape=SEbox,inner sep=0.6mm,font=\footnotesize,minimum height=4mm]

\tikzstyle{hadamard}=[fill=white,draw,inner sep=0.6mm,font=\footnotesize,minimum height=4mm,minimum width=4mm]
\tikzstyle{small hadamard}=[fill=white,draw,inner sep=0.6mm,minimum height=1.5mm,minimum width=1.5mm]
\tikzstyle{small hadamard rotate}=[small hadamard,rotate=45]
\tikzstyle{dhadamard}=[hadamard,doubled]
\tikzstyle{small dhadamard}=[small hadamard,doubled]
\tikzstyle{small dhadamard rotate}=[small hadamard rotate,doubled]
\tikzstyle{antipode}=[white dot,inner sep=0.3mm,font=\footnotesize]

\tikzstyle{scalar}=[diamond,draw,inner sep=0.5pt,font=\small]
\tikzstyle{dscalar}=[diamond,doubled, draw,inner sep=0.5pt,font=\small]

\tikzstyle{small box}=[rectangle,inline text,fill=white,draw,minimum height=5mm,yshift=-0.5mm,minimum width=5mm,font=\small]
\tikzstyle{small gray box}=[small box,fill=gray!30]
\tikzstyle{medium box}=[rectangle,inline text,fill=white,draw,minimum height=5mm,yshift=-0.5mm,minimum width=8mm,font=\small]
\tikzstyle{square box}=[small box] 
\tikzstyle{medium gray box}=[small box,fill=gray!30]
\tikzstyle{semilarge box}=[rectangle,inline text,fill=white,draw,minimum height=5mm,yshift=-0.5mm,minimum width=12.5mm,font=\small]
\tikzstyle{large box}=[rectangle,inline text,fill=white,draw,minimum height=5mm,yshift=-0.5mm,minimum width=15mm,font=\small]
\tikzstyle{large gray box}=[small box,fill=gray!30]

\tikzstyle{Bayes box}=[rectangle,fill=black,draw, minimum height=3mm, minimum width=3mm]

\tikzstyle{gray square point}=[small box,fill=gray!50]

\tikzstyle{dphase box white}=[dhadamard]
\tikzstyle{dphase box gray}=[dhadamard,fill=gray!50!white]
\tikzstyle{phase box white}=[hadamard]
\tikzstyle{phase box gray}=[hadamard,fill=gray!50!white]

\tikzstyle{point}=[regular polygon,regular polygon sides=3,draw,scale=0.75,inner sep=-0.5pt,minimum width=9mm,fill=white,regular polygon rotate=180]
\tikzstyle{copoint}=[regular polygon,regular polygon sides=3,draw,scale=0.75,inner sep=-0.5pt,minimum width=9mm,fill=white]
\tikzstyle{dpoint}=[point,doubled]
\tikzstyle{dcopoint}=[copoint,doubled]

\tikzstyle{wide copoint}=[fill=white,draw,shape=isosceles triangle,shape border rotate=90,isosceles triangle stretches=true,inner sep=0pt,minimum width=1.5cm,minimum height=6.12mm]
\tikzstyle{wide point}=[fill=white,draw,shape=isosceles triangle,shape border rotate=-90,isosceles triangle stretches=true,inner sep=0pt,minimum width=1.5cm,minimum height=6.12mm,yshift=-0.0mm]
\tikzstyle{wide point plus}=[fill=white,draw,shape=isosceles triangle,shape border rotate=-90,isosceles triangle stretches=true,inner sep=0pt,minimum width=1.74cm,minimum height=7mm,yshift=-0.0mm]

\tikzstyle{wide dpoint}=[fill=white,doubled,draw,shape=isosceles triangle,shape border rotate=-90,isosceles triangle stretches=true,inner sep=0pt,minimum width=1.5cm,minimum height=6.12mm,yshift=-0.0mm]

\tikzstyle{tinypoint}=[regular polygon,regular polygon sides=3,draw,scale=0.55,inner sep=-0.15pt,minimum width=6mm,fill=white,regular polygon rotate=180] 

\tikzstyle{white point}=[point]
\tikzstyle{white dpoint}=[dpoint]
\tikzstyle{green point}=[white point] 
\tikzstyle{white copoint}=[copoint]
\tikzstyle{gray point}=[point,fill=gray!40!white]
\tikzstyle{gray dpoint}=[gray point,doubled]
\tikzstyle{red point}=[gray point] 
\tikzstyle{gray copoint}=[copoint,fill=gray!40!white]
\tikzstyle{gray dcopoint}=[gray copoint,doubled]

\tikzstyle{white point guide}=[regular polygon,regular polygon sides=3,font=\scriptsize,draw,scale=0.65,inner sep=-0.5pt,minimum width=9mm,fill=white,regular polygon rotate=180]

\tikzstyle{black point}=[point,fill=black,font=\color{white}]
\tikzstyle{black copoint}=[copoint,fill=black,font=\color{white}]

\tikzstyle{tiny gray point}=[tinypoint,fill=gray!40!white]

\tikzstyle{diredge}=[->]
\tikzstyle{ddiredge}=[<->]
\tikzstyle{rdiredge}=[<-]
\tikzstyle{thickdiredge}=[->, very thick]
\tikzstyle{pointer edge}=[->,very thick,gray]
\tikzstyle{pointer edge part}=[very thick,gray]
\tikzstyle{dashed edge}=[dashed]
\tikzstyle{thick dashed edge}=[very thick,dashed]
\tikzstyle{thick gray dashed edge}=[thick dashed edge,gray!40]
\tikzstyle{thick map edge}=[very thick,|->]

\makeatletter
\newcommand{\boxshape}[3]{%
\pgfdeclareshape{#1}{
\inheritsavedanchors[from=rectangle] 
\inheritanchorborder[from=rectangle]
\inheritanchor[from=rectangle]{center}
\inheritanchor[from=rectangle]{north}
\inheritanchor[from=rectangle]{south}
\inheritanchor[from=rectangle]{west}
\inheritanchor[from=rectangle]{east}
\backgroundpath{
\southwest \pgf@xa=\pgf@x \pgf@ya=\pgf@y
\northeast \pgf@xb=\pgf@x \pgf@yb=\pgf@y

\@tempdima=#2
\@tempdimb=#3

\pgfpathmoveto{\pgfpoint{\pgf@xa - 5pt + \@tempdima}{\pgf@ya}}
\pgfpathlineto{\pgfpoint{\pgf@xa - 5pt - \@tempdima}{\pgf@yb}}
\pgfpathlineto{\pgfpoint{\pgf@xb + 5pt + \@tempdimb}{\pgf@yb}}
\pgfpathlineto{\pgfpoint{\pgf@xb + 5pt - \@tempdimb}{\pgf@ya}}
\pgfpathlineto{\pgfpoint{\pgf@xa - 5pt + \@tempdima}{\pgf@ya}}
\pgfpathclose
}
}}

\boxshape{NEbox}{0pt}{0pt}
\boxshape{SEbox}{0pt}{-5pt}
\boxshape{NWbox}{5pt}{0pt}
\boxshape{SWbox}{-5pt}{0pt}
\boxshape{EBox}{-3pt}{3pt}
\boxshape{WBox}{3pt}{-3pt}
\makeatother

\tikzstyle{cloud}=[shape=cloud,draw,minimum width=1.5cm,minimum height=1.5cm]

\tikzstyle{map}=[draw,shape=NEbox,inner sep=2pt,minimum height=6mm,fill=white]
\tikzstyle{dashedmap}=[draw,dashed,gray,shape=NEbox,inner sep=2pt,minimum height=6mm,fill=white]
\tikzstyle{medium dashedmap}=[draw,dashed,gray,shape=NEbox,inner sep=2pt,minimum height=6mm,fill=white,minimum width=7mm]
\tikzstyle{semilarge dashedmap}=[draw,dashed,gray,shape=NEbox,inner sep=2pt,minimum height=6mm,fill=white,minimum width=9.5mm]
\tikzstyle{large dashedmap}=[draw,dashed,gray,shape=NEbox,inner sep=2pt,minimum height=6mm,fill=white,minimum width=12mm]
\tikzstyle{very large dashedmap}=[draw,dashed,gray,shape=NEbox,inner sep=2pt,minimum height=6mm,fill=white,minimum width=17mm]

\tikzstyle{dashed map}=[fill=white, draw=gray, shape=rectangle, style=map, dashed]

\tikzstyle{mapdag}=[draw,shape=SEbox,inner sep=2pt,minimum height=6mm,fill=white]
\tikzstyle{mapadj}=[draw,shape=SEbox,inner sep=2pt,minimum height=6mm,fill=white]
\tikzstyle{maptrans}=[draw,shape=SWbox,inner sep=2pt,minimum height=6mm,fill=white]
\tikzstyle{mapconj}=[draw,shape=NWbox,inner sep=2pt,minimum height=6mm,fill=white]

\tikzstyle{medium map}=[draw,shape=NEbox,inner sep=2pt,minimum height=6mm,fill=white,minimum width=7mm]
\tikzstyle{medium map dag}=[draw,shape=SEbox,inner sep=2pt,minimum height=6mm,fill=white,minimum width=7mm]
\tikzstyle{medium map adj}=[draw,shape=SEbox,inner sep=2pt,minimum height=6mm,fill=white,minimum width=7mm]
\tikzstyle{medium map trans}=[draw,shape=SWbox,inner sep=2pt,minimum height=6mm,fill=white,minimum width=7mm]
\tikzstyle{medium map conj}=[draw,shape=NWbox,inner sep=2pt,minimum height=6mm,fill=white,minimum width=7mm]
\tikzstyle{semilarge map}=[draw,shape=NEbox,inner sep=2pt,minimum height=6mm,fill=white,minimum width=9.5mm]
\tikzstyle{semilarge map trans}=[draw,shape=SWbox,inner sep=2pt,minimum height=6mm,fill=white,minimum width=9.5mm]
\tikzstyle{semilarge map adj}=[draw,shape=SEbox,inner sep=2pt,minimum height=6mm,fill=white,minimum width=9.5mm]
\tikzstyle{semilarge map dag}=[draw,shape=SEbox,inner sep=2pt,minimum height=6mm,fill=white,minimum width=9.5mm]
\tikzstyle{semilarge map conj}=[draw,shape=NWbox,inner sep=2pt,minimum height=6mm,fill=white,minimum width=9.5mm]
\tikzstyle{large map}=[draw,shape=NEbox,inner sep=2pt,minimum height=6mm,fill=white,minimum width=12mm]
\tikzstyle{large map conj}=[draw,shape=NWbox,inner sep=2pt,minimum height=6mm,fill=white,minimum width=12mm]
\tikzstyle{very large map}=[draw,shape=NEbox,inner sep=2pt,minimum height=6mm,fill=white,minimum width=17mm]
\tikzstyle{very very large map}=[draw,shape=NEbox,inner sep=2pt,minimum height=6mm,fill=white,minimum width=50mm]
\tikzstyle{large map dag}=[draw,shape=SEbox,inner sep=2pt,minimum height=6mm,fill=white,minimum width=12mm]

\tikzstyle{medium dmap}=[draw,doubled,shape=NEbox,inner sep=2pt,minimum height=6mm,fill=white,minimum width=7mm]
\tikzstyle{medium dmap dag}=[draw,doubled,shape=SEbox,inner sep=2pt,minimum height=6mm,fill=white,minimum width=7mm]
\tikzstyle{medium dmap adj}=[draw,doubled,shape=SEbox,inner sep=2pt,minimum height=6mm,fill=white,minimum width=7mm]
\tikzstyle{medium dmap trans}=[draw,doubled,shape=SWbox,inner sep=2pt,minimum height=6mm,fill=white,minimum width=7mm]
\tikzstyle{medium dmap conj}=[draw,doubled,shape=NWbox,inner sep=2pt,minimum height=6mm,fill=white,minimum width=7mm]
\tikzstyle{semilarge dmap}=[draw,doubled,shape=NEbox,inner sep=2pt,minimum height=6mm,fill=white,minimum width=9.5mm]
\tikzstyle{semilarge dmap trans}=[draw,doubled,shape=SWbox,inner sep=2pt,minimum height=6mm,fill=white,minimum width=9.5mm]
\tikzstyle{semilarge dmap adj}=[draw,doubled,shape=SEbox,inner sep=2pt,minimum height=6mm,fill=white,minimum width=9.5mm]
\tikzstyle{semilarge dmap dag}=[draw,doubled,shape=SEbox,inner sep=2pt,minimum height=6mm,fill=white,minimum width=9.5mm]
\tikzstyle{semilarge dmap conj}=[draw,doubled,shape=NWbox,inner sep=2pt,minimum height=6mm,fill=white,minimum width=9.5mm]
\tikzstyle{large dmap}=[draw,doubled,shape=NEbox,inner sep=2pt,minimum height=6mm,fill=white,minimum width=12mm]
\tikzstyle{large dmap conj}=[draw,doubled,shape=NWbox,inner sep=2pt,minimum height=6mm,fill=white,minimum width=12mm]
\tikzstyle{large dmap trans}=[draw,doubled,shape=SWbox,inner sep=2pt,minimum height=6mm,fill=white,minimum width=12mm]
\tikzstyle{large dmap adj}=[draw,doubled,shape=SEbox,inner sep=2pt,minimum height=6mm,fill=white,minimum width=12mm]
\tikzstyle{large dmap dag}=[draw,doubled,shape=SEbox,inner sep=2pt,minimum height=6mm,fill=white,minimum width=12mm]
\tikzstyle{very large dmap}=[draw,doubled,shape=NEbox,inner sep=2pt,minimum height=6mm,fill=white,minimum width=19.5mm]

\tikzstyle{muxbox}=[draw,shape=rectangle,minimum height=3mm,minimum width=3mm,fill=white]
\tikzstyle{dmuxbox}=[muxbox,doubled]

\tikzstyle{box}=[draw,shape=rectangle,inner sep=2pt,minimum height=6mm,minimum width=6mm,fill=white]
\tikzstyle{dbox}=[draw,doubled,shape=rectangle,inner sep=2pt,minimum height=6mm,minimum width=6mm,fill=white]
\tikzstyle{dmap}=[draw,doubled,shape=NEbox,inner sep=2pt,minimum height=6mm,fill=white]
\tikzstyle{dmapdag}=[draw,doubled,shape=SEbox,inner sep=2pt,minimum height=6mm,fill=white]
\tikzstyle{dmapadj}=[draw,doubled,shape=SEbox,inner sep=2pt,minimum height=6mm,fill=white]
\tikzstyle{dmaptrans}=[draw,doubled,shape=SWbox,inner sep=2pt,minimum height=6mm,fill=white]
\tikzstyle{dmapconj}=[draw,doubled,shape=NWbox,inner sep=2pt,minimum height=6mm,fill=white]

\tikzstyle{ddmap}=[draw,doubled,dashed,shape=NEbox,inner sep=2pt,minimum height=6mm,fill=white]
\tikzstyle{ddmapdag}=[draw,doubled,dashed,shape=SEbox,inner sep=2pt,minimum height=6mm,fill=white]
\tikzstyle{ddmapadj}=[draw,doubled,dashed,shape=SEbox,inner sep=2pt,minimum height=6mm,fill=white]
\tikzstyle{ddmaptrans}=[draw,doubled,dashed,shape=SWbox,inner sep=2pt,minimum height=6mm,fill=white]
\tikzstyle{ddmapconj}=[draw,doubled,dashed,shape=NWbox,inner sep=2pt,minimum height=6mm,fill=white]

\boxshape{sNEbox}{0pt}{3pt}
\boxshape{sSEbox}{0pt}{-3pt}
\boxshape{sNWbox}{3pt}{0pt}
\boxshape{sSWbox}{-3pt}{0pt}
\tikzstyle{smap}=[draw,shape=sNEbox,fill=white]
\tikzstyle{smapdag}=[draw,shape=sSEbox,fill=white]
\tikzstyle{smapadj}=[draw,shape=sSEbox,fill=white]
\tikzstyle{smaptrans}=[draw,shape=sSWbox,fill=white]
\tikzstyle{smapconj}=[draw,shape=sNWbox,fill=white]

\tikzstyle{dsmap}=[draw,dashed,shape=sNEbox,fill=white]
\tikzstyle{dsmapdag}=[draw,dashed,shape=sSEbox,fill=white]
\tikzstyle{dsmaptrans}=[draw,dashed,shape=sSWbox,fill=white]
\tikzstyle{dsmapconj}=[draw,dashed,shape=sNWbox,fill=white]

\boxshape{mNEbox}{0pt}{10pt}
\boxshape{mSEbox}{0pt}{-10pt}
\boxshape{mNWbox}{10pt}{0pt}
\boxshape{mSWbox}{-10pt}{0pt}
\tikzstyle{mmap}=[draw,shape=mNEbox]
\tikzstyle{mmapdag}=[draw,shape=mSEbox]
\tikzstyle{mmaptrans}=[draw,shape=mSWbox]
\tikzstyle{mmapconj}=[draw,shape=mNWbox]

\tikzstyle{mmapgray}=[draw,fill=gray!40!white,shape=mNEbox]
\tikzstyle{smapgray}=[draw,fill=gray!40!white,shape=sNEbox]

\makeatletter

\pgfdeclareshape{cornerpoint}{
\inheritsavedanchors[from=rectangle] 
\inheritanchorborder[from=rectangle]
\inheritanchor[from=rectangle]{center}
\inheritanchor[from=rectangle]{north}
\inheritanchor[from=rectangle]{south}
\inheritanchor[from=rectangle]{west}
\inheritanchor[from=rectangle]{east}
\backgroundpath{
\southwest \pgf@xa=\pgf@x \pgf@ya=\pgf@y
\northeast \pgf@xb=\pgf@x \pgf@yb=\pgf@y

\pgfmathsetmacro{\pgf@shorten@left}{\pgfkeysvalueof{/tikz/shorten left}}
\pgfmathsetmacro{\pgf@shorten@right}{\pgfkeysvalueof{/tikz/shorten right}}

\pgfpathmoveto{\pgfpoint{0.5 * (\pgf@xa + \pgf@xb)}{\pgf@ya - 5pt}}
\pgfpathlineto{\pgfpoint{\pgf@xa - 8pt + \pgf@shorten@left}{\pgf@yb - 1.5 * \pgf@shorten@left}}
\pgfpathlineto{\pgfpoint{\pgf@xa - 8pt + \pgf@shorten@left}{\pgf@yb}}
\pgfpathlineto{\pgfpoint{\pgf@xb + 8pt - \pgf@shorten@right}{\pgf@yb}}
\pgfpathlineto{\pgfpoint{\pgf@xb + 8pt - \pgf@shorten@right}{\pgf@yb - 1.5 * \pgf@shorten@right}}
\pgfpathclose
}
}

\pgfdeclareshape{cornercopoint}{
\inheritsavedanchors[from=rectangle] 
\inheritanchorborder[from=rectangle]
\inheritanchor[from=rectangle]{center}
\inheritanchor[from=rectangle]{north}
\inheritanchor[from=rectangle]{south}
\inheritanchor[from=rectangle]{west}
\inheritanchor[from=rectangle]{east}
\backgroundpath{
\southwest \pgf@xa=\pgf@x \pgf@ya=\pgf@y
\northeast \pgf@xb=\pgf@x \pgf@yb=\pgf@y

\pgfmathsetmacro{\pgf@shorten@left}{\pgfkeysvalueof{/tikz/shorten left}}
\pgfmathsetmacro{\pgf@shorten@right}{\pgfkeysvalueof{/tikz/shorten right}}

\pgfpathmoveto{\pgfpoint{0.5 * (\pgf@xa + \pgf@xb)}{\pgf@yb + 5pt}}
\pgfpathlineto{\pgfpoint{\pgf@xa - 8pt + \pgf@shorten@left}{\pgf@ya + 1.5 * \pgf@shorten@left}}
\pgfpathlineto{\pgfpoint{\pgf@xa - 8pt + \pgf@shorten@left}{\pgf@ya}}
\pgfpathlineto{\pgfpoint{\pgf@xb + 8pt - \pgf@shorten@right}{\pgf@ya}}
\pgfpathlineto{\pgfpoint{\pgf@xb + 8pt - \pgf@shorten@right}{\pgf@ya + 1.5 * \pgf@shorten@right}}
\pgfpathclose
}
}

\makeatother

\pgfkeyssetvalue{/tikz/shorten left}{0pt}
\pgfkeyssetvalue{/tikz/shorten right}{0pt}

\tikzstyle{kpoint common}=[draw,fill=white,inner sep=1pt,minimum height=4mm]
\tikzstyle{kpoint sc}=[shape=cornerpoint,kpoint common]
\tikzstyle{kpoint adjoint sc}=[shape=cornercopoint,kpoint common]
\tikzstyle{kpoint}=[shape=cornerpoint,shorten left=5pt,kpoint common]
\tikzstyle{kpoint adjoint}=[shape=cornercopoint,shorten left=5pt,kpoint common]
\tikzstyle{kpoint conjugate}=[shape=cornerpoint,shorten right=5pt,kpoint common]
\tikzstyle{kpoint transpose}=[shape=cornercopoint,shorten right=5pt,kpoint common]
\tikzstyle{kpoint symm}=[shape=cornerpoint,shorten left=5pt,shorten right=5pt,kpoint common]

\tikzstyle{black kpoint}=[shape=cornerpoint,shorten left=5pt,kpoint common,fill=black,font=\color{white}]
\tikzstyle{black kpoint adjoint}=[shape=cornercopoint,shorten left=5pt,kpoint common,fill=black,font=\color{white}]
\tikzstyle{black kpointadj}=[shape=cornercopoint,shorten left=5pt,kpoint common,fill=black,font=\color{white}]

\tikzstyle{black dkpoint}=[shape=cornerpoint,shorten left=5pt,kpoint common,fill=black, doubled,font=\color{white}]
\tikzstyle{black dkpoint adjoint}=[shape=cornercopoint,shorten left=5pt,kpoint common,fill=black, doubled,font=\color{white}]
\tikzstyle{black dkpointadj}=[shape=cornercopoint,shorten left=5pt,kpoint common,fill=black, doubled,font=\color{white}] 

\tikzstyle{kpointdag}=[kpoint adjoint]
\tikzstyle{kpointadj}=[kpoint adjoint]
\tikzstyle{kpointconj}=[kpoint conjugate]
\tikzstyle{kpointtrans}=[kpoint transpose]

\tikzstyle{big kpoint}=[kpoint, minimum width=1.2 cm, minimum height=8mm, inner sep=4pt, text depth=3mm]

\tikzstyle{wide kpoint}=[kpoint, minimum width=1 cm, inner sep=2pt]
\tikzstyle{wide kpointdag}=[kpointdag, minimum width=1 cm, inner sep=2pt]
\tikzstyle{wide kpointconj}=[kpointconj, minimum width=1 cm, inner sep=2pt]
\tikzstyle{wide kpointtrans}=[kpointtrans, minimum width=1 cm, inner sep=2pt]

\tikzstyle{gray kpoint}=[kpoint,fill=gray!50!white]
\tikzstyle{gray kpointdag}=[kpointdag,fill=gray!50!white]
\tikzstyle{gray kpointadj}=[kpointadj,fill=gray!50!white]
\tikzstyle{gray kpointconj}=[kpointconj,fill=gray!50!white]
\tikzstyle{gray kpointtrans}=[kpointtrans,fill=gray!50!white]

\tikzstyle{gray dkpoint}=[kpoint,fill=gray!50!white,doubled]
\tikzstyle{gray dkpointdag}=[kpointdag,fill=gray!50!white,doubled]
\tikzstyle{gray dkpointadj}=[kpointadj,fill=gray!50!white,doubled]
\tikzstyle{gray dkpointconj}=[kpointconj,fill=gray!50!white,doubled]
\tikzstyle{gray dkpointtrans}=[kpointtrans,fill=gray!50!white,doubled]

\tikzstyle{white label}=[draw,fill=white,rectangle,inner sep=0.7 mm]
\tikzstyle{gray label}=[draw,fill=gray!50!white,rectangle,inner sep=0.7 mm]
\tikzstyle{black label}=[draw,fill=black,rectangle,inner sep=0.7 mm]

\tikzstyle{dkpoint}=[kpoint,doubled]
\tikzstyle{wide dkpoint}=[wide kpoint,doubled]
\tikzstyle{dkpointdag}=[kpoint adjoint,doubled]
\tikzstyle{wide dkpointdag}=[wide kpointdag,doubled]
\tikzstyle{dkcopoint}=[kpoint adjoint,doubled]
\tikzstyle{dkpointadj}=[kpoint adjoint,doubled]
\tikzstyle{dkpointconj}=[kpoint conjugate,doubled]
\tikzstyle{dkpointtrans}=[kpoint transpose,doubled]

\tikzstyle{kscalar}=[kpoint common, shape=EBox, inner xsep=-1pt, inner ysep=3pt,font=\small]
\tikzstyle{kscalarconj}=[kpoint common, shape=WBox, inner xsep=-1pt, inner ysep=3pt,font=\small]

\tikzstyle{spekpoint}=[kpoint sc,minimum height=5mm,inner sep=3pt]
\tikzstyle{spekcopoint}=[kpoint adjoint sc,minimum height=5mm,inner sep=3pt]

\tikzstyle{dspekpoint}=[spekpoint,doubled]
\tikzstyle{dspekcopoint}=[spekcopoint,doubled]

 \tikzstyle{discard}=[circuit ee IEC, ground,rotate=180,scale=1.5,inner sep=-2mm]
 \tikzstyle{downground}=[circuit ee IEC,thick,ground,rotate=-90,scale=1.5,inner sep=-2mm]

\tikzstyle{maxmix}=[regular polygon,regular polygon sides=3,draw=black,xscale=0.4,yscale=0.3,inner sep=-0.5pt,minimum width=10mm,fill=gray,regular polygon rotate=180]

 \tikzstyle{bigground}=[regular polygon,regular polygon sides=3,draw=gray,scale=0.50,inner sep=-0.5pt,minimum width=10mm,fill=gray]

\tikzstyle{arrs}=[-latex,font=\small,auto]
\tikzstyle{arrow plain}=[arrs]
\tikzstyle{arrow dashed}=[dashed,arrs]
\tikzstyle{arrow bold}=[very thick,arrs]
\tikzstyle{arrow hide}=[draw=white!0,-]
\tikzstyle{arrow reverse}=[latex-]
\tikzstyle{cdnode}=[]

\tikzstyle{green dashed arrow}=[green, arrow dashed]
\tikzstyle{dashed blue}=[blue, dashed]
\tikzstyle{red dashed arrow}=[red, arrow dashed]
\tikzstyle{orange arrow}=[orange, arrs]
\tikzstyle{blue arrow}=[blue, arrs]
\tikzstyle{magenta arrow}=[magenta, arrs]

\tikzstyle{black dot}=[fill=black, draw=black, shape=circle, minimum size=1.5 mm, inner sep=0]
\tikzstyle{red dot}=[fill=red, draw=red, shape=circle, minimum size=1.5 mm, inner sep=0]

\tikzstyle{black arrow}=[->, very thick]
\tikzstyle{dashed line}=[-, dashed, draw=red]
\tikzstyle{red line}=[-, draw=red]
\tikzstyle{line}=[-]
\tikzstyle{dashed black}=[-, dashed]
\tikzstyle{green dashed}=[dashed, draw=green, ->]
\tikzstyle{green arrow}=[->, draw={rgb,255: red,0; green,135; blue,0}, thick]
\tikzstyle{blue arrow}=[draw=blue, ->, thick]
\tikzstyle{dashed blue}=[-, draw=blue, dashed, thick]
\tikzstyle{blue line}=[-, draw=blue, thick]

\usepackage{physics}
\usepackage{amssymb}
\usepackage{amsmath}
\usepackage{amsfonts}
\usepackage{braket}
\usepackage{mathtools}
\usepackage{amsthm}
\usepackage{ulem}
\usepackage{bbold}
\usepackage{subcaption}
\usepackage{hyperref}
\usepackage{stmaryrd}
\usepackage{xcolor}
\usepackage{relsize}
\usepackage{epigraph}
\usepackage[disable]{todonotes}

\DeclareMathSymbol{\mlq}{\mathord}{operators}{``}
\DeclareMathSymbol{\mrq}{\mathord}{operators}{`'}

\def\be{\begin{equation}}
\def\ee{\end{equation}}
\def\ba{\begin{align}}
\def\ea{\end{align}}
\newcommand{\cat}[1]{\ensuremath{\mathbf{#1}}}

\newcommand{\Rel}{\cat{Rel}}

\newtheorem{theorem}{Theorem}[section]

\newtheorem{lemma}{Lemma}[section]

\DeclareTextFontCommand{\texttt}{\ttfamily\upshape}
\DeclareTextFontCommand{\textrm}{\rmfamily\upshape}

\newcommand{\cc}{\mathcal C}

\newcommand{\ce}{\mathcal E}

\newcommand{\ch}{\mathcal H}

\newcommand{\cl}{\mathcal L}

\renewcommand{\cp}{\mathcal P}

\newcommand{\cs}{\mathcal S}

\newcommand{\Ga}{\Gamma}
\newcommand{\la}{\lambda}
\newcommand{\La}{\Lambda}

\newcommand{\inn}{\textrm{\upshape in}}

\newcommand{\out}{\textrm{\upshape out}}
\renewcommand{\int}{\textrm{\upshape int}}

\newcommand{\aug}{\textrm{\upshape aug}}

\newcommand{\cha}{\textrm{\upshape ch}}
\newcommand{\pr}{\textrm{\upshape pr}}
\newcommand{\lost}{\textrm{\upshape lost}}
\newcommand{\send}{\textrm{\upshape send}}
\newcommand{\rec}{\textrm{\upshape rec}}

\newcommand{\ind}[1]{\mathbb{1}_{\{#1\}}}

\usepackage{float}

\begin{document}
\title{Unilateral determination of causal order in a cyclic process}

\author{Ilyass Mejdoub}
 \affiliation{Télécom Paris, Institut Polytechnique de Paris, Palaiseau, France}
\affiliation{Quriosity Team, Inria Saclay, Palaiseau, France}%
\author{Augustin Vanrietvelde}%
 \email{vanrietvelde@telecom-paris.fr}
\affiliation{Télécom Paris, Institut Polytechnique de Paris, Palaiseau, France}%
\affiliation{Quriosity Team, Inria Saclay, Palaiseau, France}%

\begin{abstract}


The recent years have seen interest into the possibility for (classical as well as quantum) causal structures that, while remaining logically consistent, feature a cyclic causal order between events, opening intriguing possibilities for new physics.
In the cyclic processes displayed so far, the global causal order is determined jointly by the operations performed by agents at each event, a feature that can be certified through the introduction of causal games and (the violation of) causal inequalities.
This raises the question of whether there exist processes in which an agent acting at a single event can \textit{unilaterally} determine her causal ordering with respect to some other events.
We answer this question in the affirmative, by introducing a process in which any party may be put in a position to pick, on her own, any other party to lie in her future.
We certify this behaviour by displaying a related causal inequality that the process allows to maximally violate.
\end{abstract}
\maketitle



\tableofcontents
\newpage

\section{Introduction}


It is usually assumed that the causal order between events should be definite and acyclic, and that a scenario featuring a cyclic causal order would necessarily involve logical inconsistencies (e.g.\ of the grandfather paradox type), making its physical existence implausible. Over the past decade, however, it has been shown -- through the development of the framework of \textit{superchannels} \cite{chiribella2013quantum}, or equivalently of \textit{process matrices} \cite{oreshkov2012quantum, araujo2017purification} -- that there exist processes that display a cyclic causal order while remaining logically consistent. This has led to various investigations of these processes' classification, properties, physical plausibility, and potential uses.

Among these exotic processes, the best-known are those involving the coherent (and possibly dynamical) control of causal order by a quantum system \cite{wechs2021quantum}. However, there also exist processes that do not fall into this category, which we will call here processes beyond quantum control \cite{oreshkov2012quantum, baumeler2014maximal, baumeler2016space, araujo2017quantum, araujo2017purification, 2017Quant...1...39A, Baumeler:2021dnu}. These allow for the violation of \textit{causal inequalities}, the analogue of Bell inequalities for causal order \cite{oreshkov2012quantum, branciard2015simplest}; this violation certifies the incompatibility of observed probability distributions with a definite causal order in a device-independent way. Interestingly, \textit{classical} processes violating causal inequalities also exist, an example being the Lugano process (also called Baumeler-Wolff or Araújo-Feix) \cite{baumeler2014maximal, baumeler2016space}.

In this context, it is natural to ask how far acausality can be taken: how causally exotic can a process be, while remaining logically consistent? A common feature of the processes beyond quantum control displayed up until now is that they can be conceived of as featuring a \textit{multilateral} determination of causal order: part of the causal order (typically, the fact that one given event lies in the other events' future) is determined by correlations between the choices of operations being performed at each event. It would seem intuitive that this `globalness' of the determination of causal order is a necessary feature of processes beyond quantum control, ensuring that their causal cyclicity does not entail logical inconsistencies.

In this paper, we show that this intuition is wrong: some processes feature \textit{unilateral} determination of (part of the) causal order. By this, we mean more precisely that for each event $A$ in the process, there exists a choice of operations being performed at other events such that the operation being performed at $A$ determines, alone, which among all other events lies in $A$'s causal future.

We first formalise unilateral determination of causal order by introducing a causal game, played collaboratively by parties acting at each event, that encapsulates this notion. We show that this game satisfies a causal inequality: a bound on the winning probability of parties acting in any fixed causal order. Using a recently proposed method for building exotic processes, we then build a new process and prove that its use allows for a deterministic winning strategy in our causal game, violating the causal inequality and demonstrating unilateral determination of causal order.

\section{A causal game for unilateral determination of causal order}

We now introduce a new causal game, aimed at pinning down the notion of unilateral determination of causal order.

Denote by $A_1,...,A_n$, with $n \geq 5$, the participants to this game. Each $A_q$ receives a binary input ${I^q \in \{0,1\}}$, picked uniformly at random, and can output a binary value $O^q \in \{0,1\} $. The referee also chooses uniformly at random a `sender' party $Q_\send \in \{1, \ldots, n\}$, and a `receiver' party $Q_\rec$, with $Q_\send \neq Q_\rec$. The label $Q_\send$ is communicated to all parties, but $Q_\rec$ is only communicated to the sender $A^{Q_\send}$. The game is won if the receiver correctly guesses the input of the sender, i.e.
\begin{equation}
    \mathbb{P}(\textrm{success}) := \mathbb{P}(O^{Q_\rec} = I^{Q_\send}) \, .
\end{equation}

This should be contrasted with the form of a causal game for the Lugano process, as presented e.g.\ in Ref.\ \cite{vanrietvelde2022}. The Lugano process can also be used to win a game in which the referee picks at random a sender and receiver, the former of which has to communicate a message to the latter. However, this would require the identity of both the sender and receiver to be communicated by the referee to all parties.

In contrast, here, only the sender's identity is communicated to everyone, while the receiver's identity is communicated to the sender alone. This makes the game starkly more difficult; in intuitive terms, a winning strategy now requires the ability for the sender to determine on her own which party will be given her input, as opposed to the identity of this party being determined jointly by the actions of all agents (as happens, for instance, in a strategy based on access to the Lugano process).

In a definite causal order scenario, the game we introduced can't be won with certainty, as we have the causal inequality
\begin{equation*}
    \mathbb{P}_{\text{DCO}}(\text{success}) \leq \dfrac{3}{4} \, ,
\end{equation*}
satisfied by all strategies featuring (possibly classical mixtures of) definite causal order. This probability can be computed as follows. If the sender happens to be in the causal past of the receiver, she transmits her input through the chain of subsequent parties and the receiver can output the correct answer. Otherwise, the receiver will have to guess the sender's input, which will be correct with probability $\dfrac{1}{2}$.
Hence the probability of success is
\be
\begin{split}
    &\mathbb{P}_{\text{DCO}}(\text{success}) \\
    &= \mathbb{P}(\text{$P_s$ before $P_r$})\mathbb{P}(\text{success}|\text{$P_s$ before $P_r$}) \\
    &\quad + \mathbb{P}(\text{$P_r$ before $P_s$})\mathbb{P}(\text{success}|\text{$P_r$ before $P_s$}) \\
    &= \dfrac{1}{2}\cdot 1 + \dfrac{1}{2}\cdot \dfrac{1}{2} \\
    &= \dfrac{3}{4} \, .
    \end{split}
\ee
By symmetry of the roles of $P_r$ and $P_s$, we have $\mathbb{P}(\text{$P_s$ before $P_r$}) = \mathbb{P}(\text{$P_r$ before $P_s$}) = \dfrac{1}{2}$.

In contrast, access to the process introduced in this paper allows the $n$ parties to win the previous game with certainty, for any $n \geq 5$.

\section{Routed circuits}
In order to specify the new process, we will employ the framework of routed circuits \cite{vanrietvelde:phd}. We very briefly and semi-formally recap the methods employed, and refer the reader to past work for the technical details. Here, a quantum system $A$ is associated to a finite-dimensional Hilbert space $\ch_A$, and a quantum channel of type $A \to B$ is a completely positive trace-preserving (CPTP) map $\cc: \cl(\ch_{A}) \to \cl(\ch_{B})$, where $\cl(\ch)$ denotes the space of linear operators on $\ch$.

We start by introducing higher-order operations. A superchannel $\cs$ (sometimes also called a supermap) is a higher-order quantum operation, which maps several quantum channels (i.e.\ completely positive trace-preserving maps) $\cc_q: A_q^\inn \to A_q^\out$, for $q = 1, \ldots, n$, to one quantum channel $\cs[(\cc_q)_q]: P \to F$.\footnote{We pass quickly on this fact for the sake of the clarity of our introductory presentation, but the condition is actually stronger because of the possibility to also have ancillary systems in the channel; see Ref.\ \cite{chiribella2013quantum} for more details.} Note that a superchannel is essentially the same object as a \textit{process matrix} \cite{oreshkov2012quantum, araujo2017purification}: superchannels and process matrices are related by the Choi-Jamiolkowski isomorphism.

A superchannel accepts as an input to its $q$-th slot any quantum channel with the correct input and output spaces. A more fine-grained description can be obtained by introducing the concept of \textit{routed} channels, i.e.\ channels which we know also satisfy \textit{sectorial constraints} about how they may map sectors of their input space to sectors of their output space \cite{vanrietvelde2020routed, vanrietvelde:phd}. In a way similar to the Einstein notation, we denote $A^k$ for a sectorised Hilbert space, i.e.\ one with a preferred (finite) decomposition into sectors, $\ch_A = \bigoplus_{k \in K} \ch_{A^k}$. For two finite sets $K$ and $L$, a relation $\lambda: K \to L$ is a function from $K$ to the powerset $\cp(L)$, or equivalently, a Boolean matrix $(\la_k^l)_{k \in K}^{l \in L}$, with $\la(k) = \{l | \la_k^l = 1\}$. Given two sectorised Hilbert spaces $A^k$ and $B^l$, we say that a CP map $\cc: \cl(\ch_A) \to \cl(\ch_B)$ \textit{follows the route $\la$} if for every $k$, it maps states without support outside of the $k$-sector to states without support outside of the $\la(k)$ sectors. We say that it is a \textit{routed quantum channel} with respect to the route $\la$ if, in addition, it is trace-preserving when restricted to act on states without support outside of the sectors corresponding to $\la$'s practical inputs, i.e.\ those values of $k$ such that $\la(k) \neq \emptyset$; we then denote $\cc: A^k \overset{\la}{\to} B^l$.

This allows to define \textit{routed superchannels}: superchannels whose inputs may be restricted to be routed channels with respect to a certain route \cite{vanrietvelde2021coherent, vanrietvelde:phd}. In recent work \cite{vanrietvelde2022consistent}, it was shown that many (and conjectured that all) unitary routed superchannels can be obtained from the specification of a \textit{routed graph}, i.e.\ a graph whose arrows are decorated with finite sets and whose nodes are decorated with relations. If a routed graph satisfies two combinatorial conditions (bi-univocality and weak loops), then it is said to be valid. As proven in Ref.\ \cite{vanrietvelde2022consistent}, any valid routed graph yields a routed superchannel that acts on channels by connecting their inputs and outputs according to the graph's connectivity. Importantly, this is true even in the presence of loops in the graph, so that this method can be used to create processes with indefinite causal order. (Note that standard superchannels can easily be obtained out of a routed superchannel created in this way, by fleshing out its slots with local monopartite superchannels.)

Using these methods, the problem of creating interesting new instances of higher-order processes is turned into a combinatorial problem, that of creating valid routed graphs: this is how we will proceed in this paper.

\section{A process with unilateral determination of the causal order}
We now introduce a process featuring a unilateral determination of the causal order. Before going into the formal presentation of the process, it is enlightening to introduce it more intuitively as a voting scheme, similarly to how the Lugano process is presented in Ref.\ \cite{vanrietvelde2022consistent}. In this scheme, $n$ agents $(A_q)_{q \in \{1,..,n\}}$ vote to pick a chancellor and a president among themselves. As with the Lugano process, the non-causal aspect lies in the fact that each party learns their status (loser, chancellor, or president), and possibly receives a message from other parties, before they cast their vote. The fact that the process is a valid one (i.e.\ is described by a superchannel) shows that this prior knowledge does \textit{not} allow the agents to generate causal loops through the voting scheme.

An agent's capacity to vote is determined by their status: a loser can vote for a chancellor among other parties, a chancellor can vote for a president among other parties, and a president cannot vote. The vote's outcomes are determined as follows, where by a `strict majority' we mean $\lceil\frac{n}{2}\rceil$ or more parties. If a strict majority of (losing) parties vote for party $A_q$ as a chancellor, then $A_q$ is chancellor. If $A_q$ is chancellor, then the party $A_{q'}$ they vote for is president, \textit{if} $A_q$ still has a strict majority without $A_{q'}$'s vote; otherwise, no party is president.

Note that it is therefore possible that there be no president, or no chancellor and no president. In addition, the president (but not the chancellor) can receive arbitrarily large messages from all other parties.

\subsection{The routed Graph}


\begin{figure*}
    \centering
    \begin{subfigure}[t]{0.5\textwidth} 
        \centering
        \begin{tikzpicture}
	\begin{pgfonlayer}{nodelayer}
		\node [style=none] (0) at (-7, 0) {$A_1$};
		\node [style=none] (1) at (-4, 0) {$A_2$};
		\node [style=none] (2) at (-1, 0) {$A_3$};
		\node [style=none] (3) at (-1, 7) {X};
		\node [style=none] (4) at (2, 0) {$A_4$};
		\node [style=none] (5) at (5, 0) {$A_5$};
		\node [style=none] (6) at (-1, 9.5) {F};
		\node [style=none] (7) at (-1, 6.25) {};
		\node [style=none] (8) at (-1.5, 6.5) {};
		\node [style=none] (9) at (-2, 6.75) {};
		\node [style=none] (10) at (-0.5, 6.5) {};
		\node [style=none] (11) at (0, 6.75) {};
		\node [style=none] (12) at (-1, 7.5) {};
		\node [style=none] (13) at (-1.5, 7.25) {};
		\node [style=none] (14) at (-6.5, 0.75) {};
		\node [style=none] (15) at (-7.25, 0.75) {};
		\node [style=none] (16) at (-4, 0.75) {};
		\node [style=none] (17) at (-1, 0.75) {};
		\node [style=none] (18) at (2, 0.75) {};
		\node [style=none] (19) at (5, 0.75) {};
		\node [style=none] (20) at (-7.5, -0.75) {};
		\node [style=none] (21) at (-4.5, -0.75) {};
		\node [style=none] (22) at (-1.5, -0.75) {};
		\node [style=none] (23) at (1.5, -0.75) {};
		\node [style=none] (24) at (4.5, -0.75) {};
		\node [style=none] (25) at (-1, 9) {};
		\node [style=none] (26) at (-8, -1.5) {};
		\node [style=none] (27) at (-5, -1.5) {};
		\node [style=none] (28) at (-2, -1.5) {};
		\node [style=none] (29) at (1, -1.5) {};
		\node [style=none] (30) at (4, -1.5) {};
		\node [style=none] (31) at (-4, 9) {};
		\node [style=none] (33) at (-1.75, 9.25) {};
		\node [style=none] (35) at (-5.25, 5.5) {\color[HTML]{008700} \scalebox{0.8}{$ l^1$}};
		\node [style=none] (40) at (-8.5, -2.25) {};
		\node [style=none] (41) at (-5.5, -2.25) {};
		\node [style=none] (42) at (-2.5, -2.25) {};
		\node [style=none] (43) at (0.5, -2.25) {};
		\node [style=none] (44) at (3.5, -2.25) {};
		\node [style=none] (45) at (-2.25, 8.5) {};
		\node [style=none] (46) at (-9, -3.25) {$\color{blue} l^1 = (l^1_{\cha}, l^1_\pr)$};
		\node [style=none] (47) at (-5.75, -3.25) {$\color{blue} l^2$};
		\node [style=none] (48) at (-2.75, -3.25) {$\color{blue} l^3$};
		\node [style=none] (49) at (0.25, -3.25) {$\color{blue} l^4$};
		\node [style=none] (50) at (3.25, -3.25) {$\color{blue} l^5$};
		\node [style=none] (51) at (-4, 9.75) {$\color{blue} l^1$};
		\node [style=none] (52) at (-8.25, 9.25) {\color{blue} \footnotesize This sector is trivial};
		\node [style=none] (53) at (-8.25, 8.5) {\color{blue} \footnotesize  when $l^1_\pr=0$};
		\node [style=none] (54) at (-5.25, 2.75) {\color[HTML]{008700} \scalebox{0.65}{$v^{1 \rightarrow 2}, \dots, v^{1 \rightarrow 5}, l^1_\pr$}};
		\node [style=none] (55) at (1.25, 8.25) {\color[HTML]{008700} \scalebox{0.8}{$(v^{q\rightarrow q'})_{q \neq q'},(l^q_\pr)_q$}};
		\node [style=none] (56) at (-8.5, 2) {};
		\node [style=none] (57) at (-6.25, 2) {};
		\node [style=none] (58) at (-11.25, 2.5) {\color[HTML]{008700} \footnotesize This sector is trivial };
		\node [style=none] (59) at (-11.25, 1.75) {\color[HTML]{008700} \footnotesize when $l^1_\pr=1$};
		\node [style=none] (62) at (7, 7.5) {};
		\node [style=none] (64) at (-6.75, -0.75) {};
		\node [style=none] (65) at (-3.75, -0.75) {};
		\node [style=none] (66) at (-0.75, -0.75) {};
		\node [style=none] (67) at (2.25, -0.75) {};
		\node [style=none] (68) at (5.25, -0.75) {};
		\node [style=none] (69) at (-1, -7) {P};
		\node [style=none] (70) at (-1.75, -7) {};
		\node [style=none] (71) at (-1.5, -6.5) {};
		\node [style=none] (72) at (-1, -6.5) {};
		\node [style=none] (73) at (-0.5, -6.5) {};
		\node [style=none] (74) at (-0.25, -7) {};
	\end{pgfonlayer}
	\begin{pgfonlayer}{edgelayer}
		\draw [style=green arrow, bend left=15] (14.center) to (9.center);
		\draw [style=green arrow, bend left=15] (16.center) to (8.center);
		\draw [style=green arrow] (17.center) to (7.center);
		\draw [style=green arrow, bend right=15] (18.center) to (10.center);
		\draw [style=green arrow, bend right=15] (19.center) to (11.center);
		\draw [style=green arrow] (12.center) to (25.center);
		\draw [style=blue arrow] (26.center) to (20.center);
		\draw [style=blue arrow] (27.center) to (21.center);
		\draw [style=blue arrow] (28.center) to (22.center);
		\draw [style=blue arrow] (29.center) to (23.center);
		\draw [style=blue arrow] (30.center) to (24.center);
		\draw [style=green arrow, bend left] (15.center) to (33.center);
		\draw [style=dashed blue] (40.center) to (26.center);
		\draw [style=dashed blue] (41.center) to (27.center);
		\draw [style=dashed blue] (42.center) to (28.center);
		\draw [style=dashed blue] (43.center) to (29.center);
		\draw [style=dashed blue] (44.center) to (30.center);
		\draw [style=blue line, bend right=15] (13.center) to (45.center);
		\draw [style=dashed blue, bend right=15] (45.center) to (31.center);
		\draw [style=green arrow] (56.center) to (57.center);
		\draw [style=black arrow, bend left=45] (70.center) to (64.center);
		\draw [style=black arrow, bend left] (71.center) to (65.center);
		\draw [style=black arrow] (72.center) to (66.center);
		\draw [style=black arrow, bend right=15] (73.center) to (67.center);
		\draw [style=black arrow, bend right] (74.center) to (68.center);
	\end{pgfonlayer}
\end{tikzpicture}
        \label{fig:process-diagram}
    \end{subfigure}%
    \hfill
    \begin{subfigure}[t]{0.4\textwidth} 
        \centering
        \vspace{-5cm} 
        {\footnotesize
        \color{blue}
        \begin{align*}
            \forall q, \quad&l^q_{\cha} = \ind{\sum_{k\neq q}v_{\cha}^{k\rightarrow q} \geq 3}  \\
            &l^q_{\pr} = \sum_{k\neq q} v_{\pr}^{k\rightarrow q}\cdot \ind{\sum_{q'\neq k,q}v_{\cha}^{q' \rightarrow k} \geq 3}  \\
        \end{align*}
        \color[HTML]{008700}
        \begin{align*}
            \forall q, \quad &\sum_{k \neq q} v_{\cha}^{q\rightarrow k} = \overline{l^{q}_{\cha}}\cdot \overline{l^q_{\pr}}  \\
            \forall q, \quad  &\sum_{k \neq q} v_{\pr}^{q\rightarrow k} = l^{q}_\cha
        \end{align*}
        }
        \label{fig:process-equations}
    \end{subfigure}
    \caption{The routed graph of the proposed process. }
    \label{fig: rooted graph}
\end{figure*}
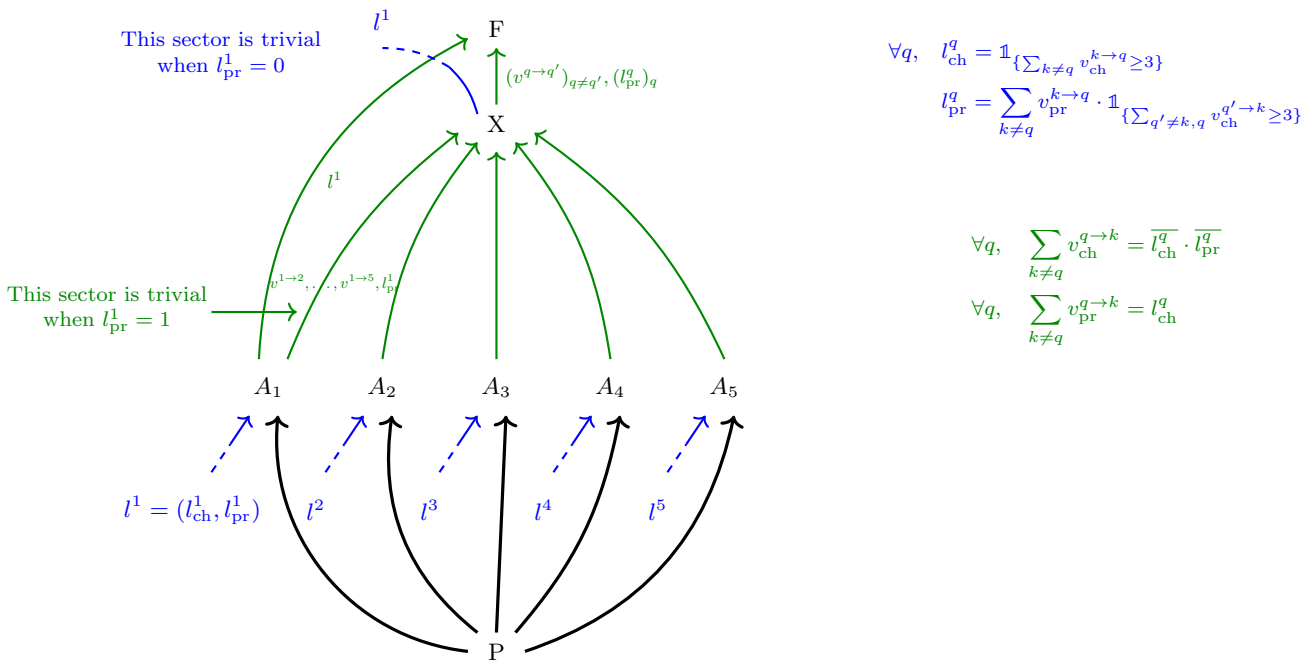

We now formally introduce the routed graph $\Ga$ from which to build our process. To simplify the presentation, we take $n=5$; generalisations to higher $n$'s are straightforward. $\Ga$ is shown in Figure \ref{fig: rooted graph}. Note that we use the shorthand method of global index constraints \cite{vanrietvelde2022consistent} to compactly specify the nodes' routes.

In this routed graph, the nodes $A_1$ to $A_5$ correspond to the voting parties, the node $X$ to a vote-counting station (analogous to the presentation in \cite{vanrietvelde2022consistent}), and the nodes $P$ and $F$ to a global past and future, respectively. Votes appear as tuples of binary values $(v^{q \rightarrow j}_{\cha}, v^{q \rightarrow j}_{\pr})_{j \neq q=1,..,5}$, with $v^{q \rightarrow j}_{\cha}$ (resp.\ $v^{q \rightarrow j}_{\pr}$) encoding whether $A_q$ votes for $A_j$ as chancellor (resp.\ as president). Statuses appear as tuples of binary values $(l^{q}_{\cha}, l^{q}_{\pr})_{q=1,..,5}$, with $l^{q}_{\cha}$ (resp.\ $l^{q}_{\pr}$) encoding whether $A_q$ is chancellor (resp.\ president). The votes $v$ and the statuses $l$ satisfy the following system of equations, where bars denote bit negation. (Note that the sum used in these equations is that of natural numbers, not of Booleans.)

\begin{subequations}
    
\label{eq: constraints}
\begin{equation}
\label{def l_ch}
    \forall q, \quad l^q_{\cha} = \ind{\sum_{k\neq q}v_{\cha}^{k\rightarrow q} \geq 3} \, ;
\end{equation}

\begin{equation}
\label{def l_q}
    \forall q, \quad l^q_{\pr} = \sum_{k\neq q} v_{\pr}^{k\rightarrow q}\cdot \ind{\sum_{q'\neq k,q}v_{\cha}^{q' \rightarrow k} \geq 3} \, ;
\end{equation}


\begin{equation}
\label{sum v_ch}
    \forall q, \quad  \sum_{k \neq q} v_{\cha}^{q\rightarrow k} = \overline{l^{q}_{\cha}}\cdot \overline{l^q_{\pr}} \, ;
\end{equation}

\begin{equation}
\label{v_p only for ch}
\forall q, \quad  \sum_{k \neq q} v_{\pr}^{q\rightarrow k} = l^{q}_\cha \, .
\end{equation}

\end{subequations}

Let us explain the intuitions behind each of these equations. (\ref{def l_ch}) enforces that a party is chancellor only if they have more than 3 votes from the other parties.
(\ref{def l_q}) states that a party $A_q$ is president if there is a vote (from the chancellor) for them and if the chancellor still holds a strict majority when only counting votes from other parties (i.e.\ excluding $A_q$'s potential vote).
(\ref{sum v_ch}) encodes the fact that only the losers of the election can vote for the chancellor, and additionally can only vote for a single candidate.
Finally, (\ref{v_p only for ch}) ensures that only the chancellor can vote for a president, and can cast a single vote.

A few handy equations which can be derived from (\ref{eq: constraints}) are proven in Appendix \ref{app: lemmas}.









\subsection{Consistency of the process}
In Appendix \ref{app: validity}, we show that $\Ga$ satisfies the two principles for validity, bi-univocality and weak loops. Therefore, by Theorem 3.1 in \cite{vanrietvelde2022consistent}, it defines a valid routed superchannel (given a specification of linear dimensions for all sectors).

\subsection{Winning the causal game}

Having proven that our new process is a valid one, we now show how it may be used to win the causal game we previously introduced. This will be natural, since, as the voting-protocol description of the process might have made clear, the process' logical structure as described by its routes precisely matches the rules of the game: it relies around the possibility for the chancellor to pick a president on her own.

In fact, we have not specified the dimensions of all sectors in the routed graph; let us start by picking a particular choice of dimensions (essentially, the simplest one) for which the corresponding routed superchannel allows to violate our causal inequality. For arrows connecting an $A_q$ to $X$, we take each allowed sector to be of dimension 1, and similarly for arrows connecting $X$ to an $A_q$ and for the arrow connecting $X$ to $F$. For arrows connecting $A_q$ to $F$, we take the $l_\pr^q = 0$ sector to be of dimension 1, and the $l_\pr^q = 1$ sector to be of dimension 4. We also take the unsectorised arrows from $P$ to each $A^q$ to be of dimension 4.

Because we are only interested in a superchannel relating the $A^q$ nodes, let us also fix a choice of input channels for other nodes. Looking at the channel $\cc_X$ for the $X$ node, taking into account that this node's route is a fully branched one with the input and output sectors in each branch being one-dimensional, we take this channel to be the identity up to bookkeping, i.e.\ the unitary channel that maps each branch's input to its unique possible corresponding output, with no dephasing between the branches. We take the channels for $P$ and $F$ to be completely depolarising ones (these channels are anyway irrelevant as far as the game is concerned). With these fixed, the superchannel turns into one with five slots $A_1, \ldots, A_5$.

These channels also follow the corresponding augmented routed of their nodes. Overall, therefore, we have fixed a channel $\cc_N$, following $\la_N^\aug$, for each node $N$.

Here again, before turning to a technical description, we first pitch at the intuitive level the way in which agents with access to this non-causal voting protocol can win our causal game with certainty. First, all losing parties in the protocol vote for the sender $A^{Q_\send}$ (whose identity they know) as a chancellor, so that the chancellor then necessarily is that sender. Having access to the bit $I^{Q_\send}$ to be sent as well as to the label $Q_\rec$ of the receiver, the sender/chancellor acts in the following way: if $I^{Q_\send} = 1$, they vote for $A^{Q_\rec}$ as president; while if $I^{Q_\send} = 0$, they purposefully vote for an arbitrary other party, say $A^{Q_\rec \oplus 1}$, where $\oplus$ is the sum modulo 5. In that way, $A^{Q_\rec}$'s presidential status is equal to $I^{Q_\send}$, so that she can simply set her output $O^{Q_\rec}$ to be equal to that status in order to win the game.\footnote{Because we used the simplest set of dimensions for our superchannel, the message sent by the sender to the receiver is encoded in whether the receiver is president. However, in instances with higher dimensions, a different strategy would be for the sender to always vote for the receiver as president and to send them their bit -- or even an arbitrary number of (qu)bits, which would e.g.\ allow to win in a more involved version of our game, -- as a message carried through the now non-trivial corresponding sector of the $A \to X$ and $X \to A$ arrows.}

We now describe a choice of specific quantum instruments $\ce_{A^q}$ for the agents that formalise the previous strategy, and can be used in order to win our causal game for unilateral determination of causal order. One can note that, although they are described in the formalism of quantum instruments, these are actually purely classical (i.e.\ decoherent) operations.

$\ce^q$, the instrument implemented by $A^q$, has classical settings corresponding $A^q$'s inputs in the game, $I^q$, $Q^q_\send$, and $Q^q_\rec$;\footnote{As per the rules of the game, we take all the $Q^q_\send$'s to be initialised by the referee to the same value $Q_\send$, and we take $Q^{Q_\send}_\rec$ to be equal to $Q_\rec$. The other $Q^q_\rec$'s are irrelevant dummy variables introduced for homogeneity of the parties' inputs; each of them could e.g.\ be initialised to a random value by the referee.} and a classical outcome that corresponds to the party's guess, $O^q$. Its inputs correspond to a Hilbert space with a preferred basis $\ket{u, (l^q_\cha, l^q_\pr)}$, where we listed first the input corresponding to the $P \to A^q$ arrow (with $u \in \{0, \ldots, 3\}$ indexing an arbitrary basis), then the input corresponding to the $X \to A^q$ arrow. Its outputs correspond to a Hilbert space with basis $\ket{((v_\cha^{q \to q'})_{q'}, (v_\pr^{q \to q'})_{q'}), l_\pr^q), (l^q_\cha, l^q_\pr, u)}$, where we listed first the outputs corresponding to the $A^q \to X$ arrow, then the outputs corresponding to the $A^q \to F$ arrow (here as well $u \in \{0, \ldots, 3\}$, with only the $u=0$ value allowed when $l_\pr^q=0$).

$\ce_{A^q}$ is specified by the following family of Kraus operators:

\begin{subequations} \label{eq: instrument}
    \be \label{eq: instrument loser} \begin{split}
        &E^{(I^q, Q^q_\send, Q^q_\rec), u, (l^q_\cha = 0, l^q_\pr =0)} \\
        &= \ket{((\delta_{q'}^{Q^q_\send})_{q'}, (0)_{q'}, 0), (0,0,u)}  \bra{u, (0, 0)} \, ;
    \end{split} \ee

    \be \label{eq: instrument chanc} \begin{split}
        &E^{(I^q, Q^q_\send, Q^q_\rec), u, (l^q_\cha = 1, l^q_\pr =0)} \\
        &= \ket{((0)_{q'}, (\delta_{q'}^{Q^q_\send \oplus \bar{I}^q})_{q'}, 0), (1,0,u)}  \bra{u, (1, 0)} \, ;
    \end{split} \ee

    \be \label{eq: instrument pres} \begin{split}
        &E^{(I^q, Q^q_\send, Q^q_\rec), u, (l^q_\cha = 0, l^q_\pr =1)} \\
        &= \ket{((0)_{q'}, (0)_{q'}, 0), (0,1,u)}  \bra{u, (0,1)} \, ;
    \end{split} \ee
\end{subequations}
furthermore, the Kraus operators in (\ref{eq: instrument loser}) and (\ref{eq: instrument chanc}) are associated with obtaining the outcome $O^q = 0$, while those in (\ref{eq: instrument pres}) are associated with obtaining the outcome $O^q = 1$.

\begin{theorem} \label{th: outcome}
    If each of the parties implements the instrument $\ce$, and the classical settings are initialised as per the game's rules, then the receiver obtains the outcome $O^{Q_\rec} = I^{Q_\send}$ with probability 1.
\end{theorem}
This Theorem is proven in Appendix \ref{app: outcome}.

\section{Conclusion}
In this paper, we showed that in causally exotic yet logically consistent processes, causal order  does not have to be determined merely by the correlations between operations at various events: we displayed a process in which a party acting at a single event may be put in a position to \textit{single-handedly} determine which among all other parties lies in her causal future. This pushes further the limits of how surprising causal structures can prove, once the acyclicity condition is dropped.

A limitation of our result is that we did not formally prove that ($N$-party generalisations of) existing cyclic processes have a bounded winning probability at our causal game for unilateral determination of causal order, although we find it very likely that it is the case. It would be important, in future work, to demonstrate more formally the separation between processes with multilateral and unilateral determinations of causal order.

It is also important to keep in mind that in our process, a party acting at a given event can unilaterally determine one among all other events to lie in her future only if the operations at other events are of a certain kind -- in intuitive terms, a party can pick someone else to be president only if other parties vote for her as a chancellor. This corresponds to the fact that in our causal game, while the identity of the receiver is only communicated to the sender, the identity of the sender is communicated to everyone. In other words, our process allows for the appearance of a causal ordering $A \to B$ with the operation at $A$ single-handedly determining $B$'s identity; but $A$ herself still has to be picked by correlations between choices of operations at other events.

Therefore, while our process constitutes a significant departure from previous instances of indefinite causal order, there remains a global aspect in its behaviour. It sounds plausible that this residual multilateralness is unavoidable in order to retain logical consistency, but further research into this question would be warranted.

\section*{Acknowledgements}
It is a pleasure to thank Nick Ormrod for his collaboration on designing a previous version of the proposed process, and Maarten Grothus for helpful criticism of that previous version. AV is supported by the STeP2 grant ANR-22-EXES-0013 of Agence Nationale de la Recherche (ANR).

\appendix
\section{Some Lemmas} \label{app: lemmas}

The following lemmas derive some implications from the satisfaction of (\ref{eq: constraints}). In each of them, we implicitly assume that we fixed a tuple of values of the $v$'s and $l$'s satisfying (\ref{eq: constraints}).
\begin{lemma}
\label{lemma: No_chan_no_pres}
If there is no chancellor, there is no president:
\begin{equation}
    \forall q, l^q_{\cha} =0 \implies \forall q', l^{q'}_{\pr} = 0 \, .
\end{equation}
\end{lemma}
\begin{proof}
    Suppose $\forall q , l^q_{\cha} = 0$. 
    By (\ref{v_p only for ch}), this entails that $\forall q, \forall k \neq q, v_\pr^{q \to k} = 0$, so by (\ref{def l_q}), $l_\pr^q = 0 \forall q$. 
    
    
\end{proof}

\begin{lemma}
    There can be at most one chancellor:
    \label{lemma: at most one chancellor}
    \begin{equation}
        \#\{q|l^q_{\cha}=1\} \leq 1 \, .
    \end{equation}
\end{lemma}
\begin{proof}
    Let us suppose that $l^{q'}_{\cha} = l^{q''}_{\cha} = 1$ with $q' \neq q''$.
    By (\ref{def l_ch}), we get that both 
    \begin{equation}
        \sum_{k\neq q'}v_{\cha}^{k \rightarrow q'} \geq 3 \text{ and } \sum_{k\neq q''}v_{\cha}^{k \rightarrow q''} \geq 3
    \end{equation} 
    are satisfied, and hence their sum is strictly greater than 5. However, by (\ref{sum v_ch}), we have that for every $q$ at most one of the $v_\cha^{q \to k}$'s can be equal to $1$, so the aforementioned sum is lesser or equal than 5, which is a contradiction.
\end{proof}

\begin{lemma}
    \label{lemma: no one in two positions}
    No one can be a chancellor and a president simultaneously.
    \begin{equation}
        \forall q, l^{q}_{\pr} \cdot l^{q}_{\cha} = 0
    \end{equation}
\end{lemma}
\begin{proof}
    Let us suppose that $\exists q', l^{q'}_{\pr} = l^{q'}_{\cha} =1$.
    From lemma \ref{lemma: at most one chancellor}, we know that no other party is chancellor: $\forall k \neq q', l^{k}_{\cha} = 0$. Hence from (\ref{v_p only for ch}), we get that $\forall k \neq q', v_{\pr}^{k\rightarrow q'} = 0$. Finally, using (\ref{def l_q}) we deduce that $l^{q'}_{\pr} = 0$ which is a contradiction.
\end{proof}

\begin{lemma}
    \label{lemma: at most one president}
    There can be at most one president:
    \begin{equation}
        \#\{q|l^q_{\pr}=1\} \leq 1 \, .
    \end{equation}
\end{lemma}
\begin{proof}
    Suppose that $l^q_\pr = 1$. By (\ref{def l_q}), there exists a $k$ such that $v_\pr^{k \to q} = 1$, so by (\ref{v_p only for ch}), $l^k_\cha = 1$ and $\forall q' \neq q, v_\pr^{k \to q'} = 0$. By Lemma \ref{lemma: at most one chancellor}, this entails that for any $k' \neq k$, $l_\cha^{k'}=0$ and thus by (\ref{v_p only for ch}) $\forall q', v_\pr^{k' \to q'} = 0$. In conclusion $v_\pr^{k' \to q'} = 1 \iff k'=k, q'=q$, so by (\ref{def l_q}), $l_\pr^{q'} = 0 \forall q' \neq q$.
\end{proof}

The rest of this section is dedicated to showing that this routed graph satisfies the two conditions for validity that, though the main theorem of \cite{vanrietvelde2022consistent}, ensure that it defines a valid process.

\section{Validity of the routed graph} \label{app: validity}

In this appendix, we formally prove that the graph is valid, i.e.\ that it satisfies the two requirements put forward in \cite{vanrietvelde2022consistent}: bi-univocality and weak loops.

\subsection{Bi-univocality}\label{section: univocality}

\subsubsection{The parties' routes}\label{subsec: party's routes}
\begin{figure}[H]
    \centering
    \resizebox{\linewidth}{!}{\begin{tikzpicture}
	\begin{pgfonlayer}{nodelayer}
		\node [style=black dot] (0) at (0, 0) {};
		\node [style=black dot] (1) at (-1.5, 5) {};
		\node [style=black dot] (2) at (1.5, 5) {};
		\node [style=black dot] (3) at (2.5, 5) {};
		\node [style=black dot] (4) at (5.5, 5) {};
		\node [style=black dot] (5) at (6.75, 5) {};
		\node [style=black dot] (6) at (4, 0) {};
		\node [style=black dot] (7) at (6.75, 0) {};
		\node [style=none] (8) at (-1.25, 4.5) {};
		\node [style=none] (9) at (-0.25, 0.5) {};
		\node [style=none] (10) at (0.25, 0.5) {};
		\node [style=none] (11) at (1.25, 4.5) {};
		\node [style=none] (12) at (0, 5) {$\dots \dots$};
		\node [style=none] (13) at (9.5, 5) {$\color{red} \dots \dots$};
		\node [style=none] (14) at (-3, 2.5) {$\lambda_{A_q}$};
		\node [style=none] (15) at (2.75, 4.5) {};
		\node [style=none] (16) at (4, 5) {$\dots \dots$};
		\node [style=none] (17) at (9.5, -1) {$\color{red}(1,1)$};
		\node [style=none] (18) at (5.25, 4.5) {};
		\node [style=none] (19) at (3.75, 0.5) {};
		\node [style=none] (20) at (4.25, 0.5) {};
		\node [style=none] (21) at (6.75, 4.5) {};
		\node [style=none] (22) at (6.75, 0.5) {};
		\node [style=none] (23) at (0, -1) {$(0,0)$};
		\node [style=none] (24) at (4, -1) {$(1,0)$};
		\node [style=none] (25) at (6.75, -1) {$(0,1)$};
		\node [style=red dot] (26) at (9.5, 0) {};
		\node [style=red dot] (28) at (8, 5) {};
		\node [style=red dot] (29) at (11, 5) {};
		\node [style=none] (30) at (7.5, 5) {};
		\node [style=none] (31) at (11.5, 5) {};
		\node [style=none] (32) at (9, 0) {};
		\node [style=none] (33) at (12.25, 2.5) {};
		\node [style=none] (34) at (10, 0) {};
		\node [style=none] (35) at (8, 5.5) {};
		\node [style=none] (36) at (11, 5.5) {};
		\node [style=none] (37) at (11.25, 2.5) {};
		\node [style=none] (40) at (15, 2.5) {\color{red} \scriptsize Outside of pratical};
		\node [style=none] (41) at (15.25, 2) {\color{red} \scriptsize input and output values};
		\node [style=none] (42) at (-5, 2.75) {};
	\end{pgfonlayer}
	\begin{pgfonlayer}{edgelayer}
		\draw [style=black arrow] (9.center) to (8.center);
		\draw [style=black arrow] (10.center) to (11.center);
		\draw [style=black arrow] (19.center) to (15.center);
		\draw [style=black arrow] (20.center) to (18.center);
		\draw [style=black arrow] (22.center) to (21.center);
		\draw [style=dashed line] (35.center) to (36.center);
		\draw [style=dashed line, bend left, looseness=1.75] (36.center) to (31.center);
		\draw [style=dashed line] (31.center) to (34.center);
		\draw [style=dashed line, bend right=60, looseness=2.00] (32.center) to (34.center);
		\draw [style=dashed line] (32.center) to (30.center);
		\draw [style=dashed line, bend left, looseness=1.50] (30.center) to (35.center);
		\draw [style=red line] (37.center) to (33.center);
	\end{pgfonlayer}
\end{tikzpicture}}
    \caption{The branched route $\lambda_{A_q}$}
     \label{fig: parties route}
\end{figure}
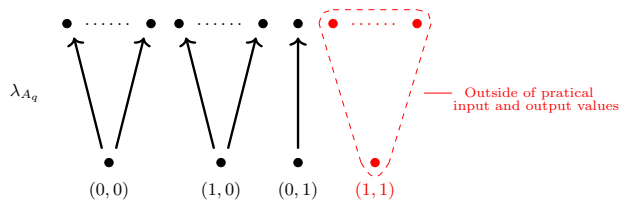
In order to check bi-univocality, we need to first describe in more details each of the nodes' route, as well as the corresponding augmented version. We start with the parties nodes $(A_q)_{q=1,..,5}$; given the obvious symmetry of the process, they all have the same route. Let us describe it for an arbitrary $q$.

The inputs of $A_q$ are indexed with $l^q:= (l_\cha^q, l_\pr^q)$, while its outputs are indexed with $l^q$ as well as $v^q := (v^{q \to k}_\cha, v^{q \to k}_\pr)_{k \neq q}$; its route is thus a branched one, with its branches labelled by the pair $(l_\cha^q, l_\pr^q)$. Since the value $(1,1)$ is not allowed by Lemma \ref{lemma: no one in two positions}, it thus has 3 branches. Furthermore,  (\ref{sum v_ch}) and (\ref{v_p only for ch}) entail that in any allowed tuple of values in which $(l_\cha^q, l_\pr^q) = (0,1)$, all the values in $v^q$ are null; so the $(0, 1)$-branch has a single output value -- this corresponds to the fact that the president cannot vote. We end up with the route $\la_{A_q}$ depicted in Figure \ref{fig: parties route}, displaying three branches, with two of them featuring bifurcations.

Furthermore, in any allowed tuple of values in which $(l_\cha^q, l_\pr^q) = (0,0)$, (\ref{sum v_ch}) and (\ref{v_p only for ch}) entail that a single $v_\cha^{q \to k}$ is equal to $1$ and the rest of the values in $v^q$ are equal to $0$; while for allowed tuple of values in which $(l_\cha^q, l_\pr^q) = (1, 0)$, they entail that a single $v_\pr^{q \to k}$ is equal to $1$ and the rest of the values in $v^q$ are equal to $0$. In other words, bifurcation choices in the $(0,0)$-branch amount to the choice of the value $k \neq q$ such that $v_\cha^{q \to k} = 1$, and bifurcation choices in the $(1,0)$-branch amount to the choice of the value $k \neq q$ such that $v_\pr^{q \to k} = 1$.

The augmented route $\lambda_{A_q}^{\aug}$ can thus be written as accepting two additional inputs $i^q_{\lost}$ and $ i^q_{\cha}$, both taking value among the $k \neq q$; and yielding three additional binary outputs $j^q_{\lost}, j^q_{\cha}, j^q_{\pr}$, each indicating whether the corresponding branch happened (with $\lost$, $\cha$ and $\pr$ being the $(0,0)$, $(1,0)$ and $(0,1)$ branch, respectively). Denoting $i^q := (i^q_{\lost}, i^q_{\cha})$ and $j^q := (j^q_{\lost}, j^q_{\cha}, j^q_{\pr})$, $\la_{A^q}^\aug$ is then the partial function from values of $(l^q, i^q)$ to values of $(l^q, v^q, j^q)$ defined by

\begin{subequations} \label{eq: parties augmented routes}
    \be \begin{split}\label{eq: parties augmented routes lost}
         &\la_{A^q}^\aug \Big( (0,0), (k, k') \Big) \\
         &\quad = \Big( (0,0), ((\delta_{q'}^k)_{q' \neq q}, (0)_{q' \neq q}) ,(1,0,0) \Big) \, ,
    \end{split} \ee
    \be \begin{split}\label{eq: parties augmented routes cha}
         &\la_{A^q}^\aug \Big( (1,0), (k, k') \Big) \\
         &\quad = \Big( (1,0), ((0)_{q' \neq q}, (\delta_{q'}^{k'})_{q' \neq q}) ,(0,1,0) \Big) \, ,
    \end{split} \ee
    \be \begin{split}\label{eq: parties augmented routes pr}
         &\la_{A^q}^\aug \Big( (0,1), (k, k') \Big) \\
         &\quad = \Big( (1,0), ((0)_{q' \neq q}, (0)_{q' \neq q}) ,(0,0,1) \Big) \, ,
    \end{split} \ee
\end{subequations}
and undefined whenever $(l_\cha^q, l_\pr^q) = (1,1)$.



\subsubsection{The routes at other nodes}\label{subsec: counting stations}
We turn to the route $\la_X$ at the counting station $X$. Since $X$ has all of the graph's indices $(v, l)$ present both in its inputs and its outputs, $\la_X$ is a branched route with no bifurcations (i.e.\ a partial function) whose branches correspond to the allowed tuples of $(v, l)$. Since it features no bifurcation, its augmented form features no extra inputs, and one extra binary output for each allowed value of $(v, l)$, recording whether this value was instantiated.

The node $P$ has no indices in its inputs or outputs and thus has a trivial route. The node $F$, featuring no indices in its outputs, has a route with a singleton codomain; this route therefore features a single branch with no bifurcations, and is equal to its augmented form.



\subsubsection{The choice relation}
The choice relation $\La$, defined as the composition of the augmented routes of all nodes, takes as input the bifurcation choices  $\vec{i} := (i^q_{\lost},i_{\cha}^q)_{q}$, and outputs $\vec{j} := (j^q)_{q} = (j^q_{\lost}, j^q_{\cha}, j^q_{\pr})_{q}$, the vector of branch statuses for the $A_q$ nodes, as well as  binary outputs $\vec{z} = (z^{(v,l)})_{(v,l)}$ labelled with the allowed values of $(v,l)$ ($\vec{z}$ corresponds to the extra outputs of $\la_X^\aug$). Univocality is satisfied if $\La$ is a function. 



For a given $\vec{i}$ and $\vec{j}$, $\La_{\vec{i}}^{\vec{j}, \vec{z}}$ is equal to $1$ if and only if there exists an assignment of values for the internal indices $(v^{q \rightarrow j}_{\cha}, v^{q \rightarrow j}_{\pr})_{j \neq q=1,..,5}$ and $(l^{q}_{\cha}, l^{q}_{\pr})_{q=1,..,5}$ with respect to which all of the augmented routes' Boolean tensors are equal to $1$, and $\La$ is a function if for any given $\vec{i}$, there is a unique $\vec{j}$ such that $\La_{\vec{i}}^{\vec{j}, \vec{z}} = 1$. The set of vectors $\vec{i}$ will be denoted as $I$, that of $\vec{j}$'s as $J$, and that of $\vec{z}$'s as $Z$. Our proof strategy relies on partitioning $I$ into sets corresponding to the scenarios `Everyone lost', `$A_q$ is the chancellor but there is no president', and `$A_q$ is the chancellor and $A_{q'}$ is the president'.


\begin{lemma}
\label{lemma: lost}
    Let 
    \be
        I_{\lost} := \Big\{\vec{i}\,\, \Big| \,\, \forall q, \# \{q'|i_{\lost}^{q'}=q \} < 3 \Big\} \,;
    \ee
    for any given $\vec{i} \in I_{\lost}$, there exists a unique $\vec{j} \in J$, $\vec{z} \in Z$ such that $\La_{\vec{i}}^{\vec{j}, \vec{z}} = 1$.
\end{lemma}

\begin{proof}
Let us fix $\vec{i} \in I_{\lost}$, $\vec{j} \in J$, $\vec{z} \in Z$ such that $\Lambda_{\vec{i}}^{\vec{j}} = 1$. We can then fix an assignment of values for the internal indices $(v^{q \rightarrow j}_{\cha}, v^{q \rightarrow j}_{\pr})_{j \neq q=1,..,5}$ and $(l^{q}_{\cha}, l^{q}_{\pr})_{q=1,..,5}$ with respect to which all of the augmented routes' coefficients are equal to $1$. Note that this value assignment therefore satisfies (\ref{eq: constraints}) in particular.

Let us fix a $q$. From (\ref{eq: parties augmented routes}), we can deduce that for any $q' \neq q$, $v_\cha^{q' \to q} \leq \delta_{q'}^{i^{q}_\lost}$, since it is equal either to $\delta_{q'}^{i^{q}_\lost}$ or to $0$. Thus we have $\sum_{q' \neq q} v_\cha^{q' \to q} \leq \sum_{q' \neq q} \delta_{q'}^{i^{q}_\lost} = \# \{q'|i_{\lost}^{q}=q' \} < 3$ by the assumption $\vec{i}\in I_\lost$. Therefore, by (\ref{def l_ch}), $l_\cha^q = 0 \forall q$. By Lemma \ref{lemma: No_chan_no_pres}, this also implies $l_\pr^q = 0 \forall q$.

Thus, for any $q$ we have $(l_\cha^q, l_\pr^q) = (0,0)$. All internal indices as well as the $\vec{j}$'s are therefore fixed univocally by the values of the $\vec{i}$'s through (\ref{eq: parties augmented routes lost}), with for every $q \neq q'$, $(j_\lost^q, j_\cha^q, j_\pr^q) = (1,0,0)$ and $(v_\cha^{q \to q'}, v_\pr^{q \to q'}) = (\delta_{q'}^{i^{q}_\lost}, 0)$. Since the values of the internal indices are fixed, this fixes the values of the $\vec{z}$'s through $\la_X^\aug$. We have thus proven that $\vec{j}$ and $\vec{z}$ have a unique possible value given $\vec{i}$.

\end{proof}


\begin{lemma}
\label{lemma: ch}
    Let 
    \begin{equation}
    \begin{split}
        I^{q}_{\cha} := &\Big\{\vec{i} \,\, \Big| \,\, \#\{q'\neq q| i^{q'}_{\lost}=q\} = 3 \\
        &\quad \textrm{ and } i^{i^q_\cha}_{\lost}=q \Big\};
    \end{split}
    \end{equation}
for any given $\vec{i} \in I^q_{\cha}$, there exists a unique $\vec{j} \in J$, $\vec{z} \in Z$ such that $\La_{\vec{i}}^{\vec{j}, \vec{z}} = 1$.
\end{lemma}

\begin{proof}
As in the previous lemma, we fix $\vec{i} \in I^q_{\cha}$, $\vec{j} \in J$, $\vec{z} \in Z$ such that $\Lambda_{\vec{i}}^{\vec{j}} = 1$, and a corresponding assignment of values for the internal indices.

Since $ \#\{q'\neq q| i^{q'}_{\lost}=q\} = 3$, we have that for $q' \neq q$, $ \#\{q''\neq q'| i^{q''}_{\lost}=q'\} < 3$. Thus, by the same reasoning as in the second paragraph of the previous lemma's proof, we can deduce that $l_\cha^{q'}=0 \forall q' \neq q$. From (\ref{eq: parties augmented routes}), this entails that $v_\pr^{q' \to q''} = 0 \forall q' \neq q, \forall q'' \neq q'$. Therefore, for any $q'' \neq q$, the sum in (\ref{def l_q}) reduces to $l^{q''}_\pr = v_\pr^{q \to q''} \cdot  \ind{\sum_{q'\neq q, q''}v_{\cha}^{q'\rightarrow q} \geq 3}$.

If for a given $q'' \neq q$ we suppose $l^{q''}_\pr = 1$, we thus necessarily have $v_\pr^{q \to q''} = 1$, which by (\ref{eq: parties augmented routes}) is only possible if $l_\cha^q = 1$ and $i_\cha^q = q''$. By definition of $I_\cha^q$, this entails that $i_\lost^{q''} = q$ and thus  $\sum_{q' \neq q, q''} v_\cha^{q' \to q} \leq \sum_{q' \neq q, q''} \delta_{q'}^{i^{q}_\lost} = \# \{q' \neq q''|i_{\lost}^{q}=q' \} = 2$; thus $\ind{\sum_{q'\neq q, q''}v_{\cha}^{q'\rightarrow q} \geq 3} = 0$, which contradicts $l^{q''}_\pr = 1$ by the equation at the end of the previous paragraph. We conclude that  $l^{q''}_\pr = 0 \forall q'' \neq q$.

We thus have that for any $q' \neq q$, $(l_\cha^{q'}, l_\pr^{q'}) = (0,0)$. This entails, through (\ref{eq: parties augmented routes lost}), that $\forall q' \neq q, v^{q' \to q}_\cha = \delta_{q}^{i^{q'}_\lost}$, so that $\sum_{q'\neq q} v_{\cha}^{q'\rightarrow q} = \sum_{q'\neq q} \delta_{q}^{i^{q'}_\lost} = \# \{q' \neq q | i^{q'}_\lost = q \} = 3$, which by (\ref{def l_ch}) implies $l_\cha^q = 1$ (and therefore $l_\pr^q = 0$ by Lemma \ref{lemma: no one in two positions}).

Thus, the $l$'s are all fixed by the value of $\vec{i}$, so that all internal values as well as the $\vec{j}$'s are fixed through (\ref{eq: parties augmented routes}), and in turn this fixes the values of $\vec{z}$ through $\la_X^\aug$.
\end{proof}

\begin{lemma}
\label{lemma: p}
Let
\be
\begin{split}
    I^{q,q'}_{\pr} &= { \{\vec{i}|\#\{k\neq q,q'|i^{k}_{\lost}=q\} \geq 3 \text{ and } i^q_{\cha} = q'\} } \, ;
\end{split}
\ee
for any given $\vec{i} \in I_{\cha}$, there exists a unique $\vec{j} \in J$, $\vec{z} \in Z$ such that $\La_{\vec{i}}^{\vec{j}, \vec{z}} = 1$. $\vec{j}$ satisfies $j^{k}_\pr = \delta^k_{q'}$ .
\end{lemma}
\begin{proof}

As in the previous lemmas, we fix $\vec{i} \in I^q_{\cha}$, $\vec{j} \in J$, $\vec{z} \in Z$ such that $\Lambda_{\vec{i}}^{\vec{j}} = 1$, and a corresponding assignment of values for the internal indices.

By the same reasoning as in the previous lemma's proof, we can deduce that for $k \neq q$, $\#\{q''\neq k|i^{q''}_{\lost}=k\} < 3$ and thus that $l^k_\cha = 0$ and (through (\ref{eq: parties augmented routes})) $v^{k \to k'}_{\pr} = 0 \forall k' \neq k$ . Furthermore, $i_q^\cha = q'$ entails through (\ref{eq: parties augmented routes}) that $v^{q \to k'}_\pr =0 \forall k' \neq q, q'$. Thus, (\ref{def l_q}) yields $l^{k'}_\pr = 0 \forall k' \neq q, q'$. Therefore, for $k \neq q, q'$, $(l^k_\cha, l^k_\pr) = (0,0)$. Through (\ref{eq: parties augmented routes lost}), this entails that for $k \neq q, q'$ and $k' \neq k$, $v_\cha^{k \to k'} = \delta_{k'}^{i^k_\lost}$. Therefore, $\sum_{k \neq q} v_\cha^{k \to q} \geq \sum_{k \neq q, q'} v_\cha^{k \to q} = \sum_{k \neq q, q'} \delta_{k}^{i^{q}_\lost} = \# \{k \neq q, q'|i_{\lost}^{k}=q \} \geq 3$, so by (\ref{def l_q}), $l_\cha^q = 1$ (and thus through Lemma \ref{lemma: no one in two positions}, $l^q_{\pr} = 0$).

By (\ref{eq: parties augmented routes cha}), this entails $v^{q \to q'}_\pr = \delta_{q'}^{i_\cha^q} = 1$, which together with $\sum_{k \neq q, q'} v_\cha^{k \to q} = \sum_{k \neq q, q'} \delta_{k}^{i^{q}_\lost} = \# \{k \neq q, q'|i_{\lost}^{k}=q \} \geq 3$ implies, by (\ref{def l_q}), that $l^{q'}_\pr = 1$ (and thus through Lemma \ref{lemma: no one in two positions}, $l^{q'}_{\cha} = 0$.)

As in the previous lemmas, we conclude that the $l$'s are all fixed by the value of $\vec{i}$, so that all internal values as well as the $\vec{j}$'s are fixed through (\ref{eq: parties augmented routes}), and in turn this fixes the values of $\vec{z}$ through $\la_X^\aug$. In particular, for every $k$, (\ref{eq: parties augmented routes}) leads to $j^{k} = l^{k}_\pr = \delta_{q'}^{k}$.
\end{proof}

\begin{theorem}\label{thm: choice function}
    $\Lambda$ is a function.
\end{theorem}
\begin{proof}
    It is direct to see that
    \begin{equation}
        I = I_\lost \sqcup \left( \bigsqcup_{q} I^{q}_{\cha} \right) \bigsqcup 
        \left(\bigsqcup_{\substack{q,q' \\ q\neq q'}} I^{q,q'}_{\pr}\right)
    \end{equation}
so that the previous lemmas prove this theorem by case disjunction.
\end{proof}

\subsubsection{Co-univocality}
Proving univocality of the reverse routed graph $\Ga^\top$ is easier: in $\Ga^\top$, the only bifurcation is the one in $F$'s unique branch. Furthermore, since all of $\Ga$'s indices are present in $F$'s inputs, picking any given (reverse) bifurcation choice at this node fixes all indices in the graph, and therefore all branches, so $\Ga^\top$'s choice relation is a function. 


\subsection{Weak loops}
\label{section: weak loops}

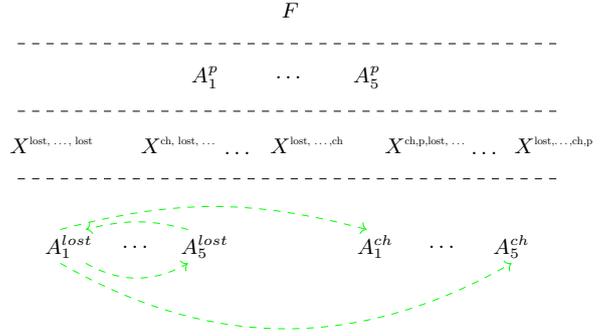
\begin{figure}
    \centering
    \resizebox{0.9\linewidth}{!}{\begin{tikzpicture}
	\begin{pgfonlayer}{nodelayer}
		\node [style=none] (2) at (-9, 2) {};
		\node [style=none] (3) at (7, 2) {};
		\node [style=none] (4) at (-9, 4) {};
		\node [style=none] (5) at (7, 4) {};
		\node [style=none] (6) at (-9, 6) {};
		\node [style=none] (7) at (7, 6) {};
		\node [style=none] (8) at (-1, 7) {$F$};
		\node [style=none] (9) at (-3.5, 5) {$A_1^p$};
		\node [style=none] (10) at (1.25, 5) {$A_5^p$};
		\node [style=none] (11) at (-1, 5) {$\dots$};
		\node [style=none] (12) at (-8, 3) {$X^{\scalebox{0.5}{lost, \dots, lost}}$};
		\node [style=none] (13) at (-4.25, 3) {$X^{\scalebox{0.5}{ch, lost, \dots}}$};
		\node [style=none] (14) at (-2.5, 2.75) {$\dots$};
		\node [style=none] (15) at (-0.5, 3) {$X^{\scalebox{0.5}{lost, \dots ,ch}}$};
		\node [style=none] (16) at (6.75, 3) {$X^{\scalebox{0.5}{lost,\dots,ch,p}}$};
		\node [style=none] (17) at (3, 3) {$X^{\scalebox{0.5}{ch,p,lost, \dots}}$};
		\node [style=none] (18) at (4.75, 2.75) {$\dots$};
		\node [style=none] (19) at (-7, -0.5) {};
		\node [style=none] (20) at (-4, 0.5) {};
		\node [style=none] (21) at (-4, -0.5) {};
		\node [style=none] (22) at (5.5, 0) {$A_5^{ch}$};
		\node [style=none] (23) at (1.5, 0) {$A_1^{ch}$};
		\node [style=none] (24) at (3.5, 0) {$\dots$};
		\node [style=none] (25) at (-7.5, 0) {$A_1^{lost}$};
		\node [style=none] (26) at (-5.5, 0) {$\dots$};
		\node [style=none] (27) at (-3.5, 0) {$A_5^{lost}$};
		\node [style=none] (28) at (-7.75, 0.5) {};
		\node [style=none] (29) at (5.5, 0.5) {};
		\node [style=none] (30) at (1.25, 0.5) {};
		\node [style=none] (31) at (-7, 0.5) {};
		\node [style=none] (32) at (1.25, -0.5) {};
		\node [style=none] (33) at (1.75, -0.5) {};
		\node [style=none] (34) at (5.5, -0.5) {};
		\node [style=none] (35) at (-7.75, -0.5) {};
	\end{pgfonlayer}
	\begin{pgfonlayer}{edgelayer}
		\draw [style=dashed black] (2.center) to (3.center);
		\draw [style=dashed black] (4.center) to (5.center);
		\draw [style=dashed black] (6.center) to (7.center);
		\draw [style=green dashed, bend right] (19.center) to (21.center);
		\draw [style=green dashed, bend left=15] (28.center) to (30.center);
		\draw [style=green dashed, bend right] (35.center) to (34.center);
		\draw [style=green dashed, bend left=345] (20.center) to (31.center);
	\end{pgfonlayer}
\end{tikzpicture}}
    \caption{A simplified version of the branch graph, only certain green arrows have been represented.}
    \label{fig: branch graph}
\end{figure}

Checking weak loops, the second condition for validity, requires to inspect $\Ga$'s branch graph, as defined in Ref.\ \cite{vanrietvelde2022consistent}.
        


        

We will not display the whole branch graph, since to prove the weak loops principle it is sufficient to show that it only features loops of dashed arrows of the same colour. In figure \ref{fig: branch graph} we arrange the nodes in layers, and our claim is that solid and red dashed arrows only go from one layer to a higher one, while green arrows go either to the same layer or to a higher one, which entails in particular that all loops in the branch graph can only be made of green dashed arrows and are therefore weak.

This claim is proven by the following lemmas (together with obvious absences of solid arrows when there are no arrows between the corresponding nodes in the routed graph).

\begin{lemma}\label{lemma: No solid arrow from X to bottom}
    There are no solid arrows from a branch $X^{(v,l)}$ to a branch $A^q_{\lost}$ or to a branch $A^q_{ch}$.
\end{lemma}

\begin{proof}
    Let us first show that there is no solid arrow from a given $X^{(v,l)}$ to a given $A^q_{\lost}$. A first case is that there is no consistent assignment of values to indices in the graph such that both branches occur simultaneously; there is then no solid arrow $X^{(v,l)} \to A^q_{\lost}$. We therefore suppose that there exist consistent value assignments compatible with the two branches, and fix one such assignment. By definition of $A^q_\lost$, it satisfies $(l^q_{\cha}, l^q_{\pr}) = (0,0)$.  This fixes that the $X \to A_q$ arrow to a single sector, which is one-dimensional since $l^q_{\pr} = 0$, so there is no solid arrow $X^{(v,l)} \to A^q_\lost$.

    The case of $A_\cha^\lost$ is analogous: a consistent assignment of value in in which the branch $A_q^{\cha}$ happens satisfies $(l^q_{\cha}, l^q_{\pr})= (1,0)$, which again fixes the $X \to A_q$ arrow to a single sector of dimension one.
\end{proof}

\begin{lemma}\label{lemma: No solid arrow from top to X}
    There are no solid arrows from a branch $A^q_\pr$ to a branch $X^{(v,l)}$.
\end{lemma}
\begin{proof}
    Let us suppose that there exists a consistent assignment of values, compatible with $A^q_\pr$ and $X^{(v,l)}$ happening. It is then unique since $X^{(v,l)}$ happening fixes all the indices $v$ and $l$ in the graph. In particular, this fixes the $A_q \to X$ arrow to a unique sector, and furthermore that sector is one-dimensional since, by compatibility with $A^q_\pr$, $l^q_\pr = 1$.
\end{proof}

\begin{lemma}
    All green dashed arrows in the branch graph start from the $A^q_\lost$ and $A^q_\cha$ branches, and all red dashed arrows go to the $F$ branch.
\end{lemma}

\begin{proof}
    Only the $A^q_\lost$'s and $A^q_\cha$'s feature bifurcation choices, and in the reverse graph, only the unique branch of $F$ features bifurcation choices.
\end{proof}

\section{Proof of Theorem \ref{th: outcome}} \label{app: outcome}

Our proof strategy is to obtain the instrument $\ce$ out of the implementation of a more involved channel $\cc$, with the latter precisely matching the behaviour of the augmented relations of the nodes, which are specified in Appendix \ref{section: univocality}. This allows us to use the theory of routed circuits to directly derive the behaviour of the image of the $\cc$'s under $\cs_\Ga$, and then to infer from this the statistics obtained by the agents if each of them implements $\ce$.

We consider $A^q$ implementing a channel $\cc$ with auxiliary quantum inputs and outputs. More precisely, it has an auxiliary input with a preferred basis indexed by $i^q = (i^q_\lost, i^q_\cha)$ (each taking values in the $q' \neq q$); and an auxiliary output with a preferred basis indexed by $j^q = (j^q_{\lost}, j^q_{\cha}, j^q_{\pr})$ (each taking a binary value). Its non-auxiliary inputs correspond to a Hilbert space with basis $\ket{u, (l^q_\cha, l^q_\pr)}$, where we listed first the input corresponding to the $P \to A^q$ arrow (with $u \in \{0, \ldots, 3\}$ indexing an arbitrary basis), then the input corresponding to the $X \to A^q$ arrow. Its non-auxiliary outputs correspond to a Hilbert space with basis $\ket{((v_\cha^{q \to q'})_{q'}, (v_\pr^{q \to q'})_{q'}), l_\pr^q), (l^q_\cha, l^q_\pr, u)}$, where we listed first the ancillary output, then the outputs corresponding to the $A^q \to X$ arrow, then the outputs corresponding to the $A^q \to F$ arrow (here as well $u \in \{0, \ldots, 3\}$, with only the $u=0$ value allowed when $l_\pr^q=0$).

$\cc_{A^q}$ is specified by the following family of Kraus operators:

\begin{subequations}
    \be \begin{split}
        &C^{(i^q_\lost, i^q_\cha), u, (l^q_\cha = 0, l^q_\pr =0)} \\
        &= \ket{0, ((\delta_{q'}^{i^q_\cha})_{q'}, (0)_{q'}, 0), (0,0,0)} \\
        &\quad \bra{(i^q_\lost, i^q_\cha), u, (0,0)} \, ;
    \end{split} \ee

    \be \begin{split}
        &C^{(i^q_\lost, i^q_\cha), u, (l^q_\cha = 1, l^q_\pr =0)} \\
        &= \ket{0, ((0)_{q'}, (\delta_{q'}^{i^q_\pr})_{q'}, 0), (0,1,0)} \\
        &\quad \bra{(i^q_\lost, i^q_\cha), u, (1,0)} \, ;
    \end{split} \ee

    \be \begin{split}
        &C^{(i^q_\lost, i^q_\cha), u, (l^q_\cha = 0, l^q_\pr =1)} \\
        &= \ket{0, ((0)_{q'}, (0)_{q'}, 0), (0,1,u)} \\
        &\quad \bra{(i^q_\lost, i^q_\cha), u, (0,1)} \, .
    \end{split} \ee
\end{subequations}
Remembering that we also specified channels $\cc_P$, $\cc_X$ and $\cc_F$, this yields a family of channels $(\cc_N)_N$ for all nodes $N$ of $\Ga$.

$\mathcal{S}_{\Gamma}[(\cc_N)_N]$, the action of the skeletal superchannel $\cs_\Ga$ on the channels $\cc_N$, is then given by taking the tensor product of the $\cc_N$'s then tracing it over the spaces corresponding to each arrow in the graph (in the sense of the trace in the theory of CP maps). $\mathcal{S}_{\Gamma}[(\cc_N)_N]$'s input space the tensor product of the auxiliary input spaces of all $\cc_{A^q}$'s: it admits a basis labelled with $\vec{i} = (i^q)_q$. Similarly, its output space is the tensor product of the auxiliary output spaces of all $\cc_{A^q}$'s: it admits a basis labelled with $\vec{j} = (j^q)_q$.

With respect to the natural sectorisation of its inputs and outputs corresponding to the preferred bases, $\cc_{A^q}$ follows the route $\la_{A_q}^\aug$ (this can be shown using Theorem 6 in \cite{vanrietvelde2022consistent}). Furthermore, the same is true for the channels $\cc_X$, $\cc_P$ and $\cc_F$, which all follow the augmented routed associated to the corresponding node.\footnote{Strictly speaking, $\cc_X$ does not follow the route $\la^\aug_X$, as it does not feature the corresponding auxiliary outputs recording which branch of $X$ happened. However, if we define $\Tilde{\cc}_X$, featuring these outputs, and acting by recording which branch happened in them, then $\cc_X$ is obtained by tracing out $\Tilde{\cc}_X$'s auxiliary outputs. The following reasoning can then formally be followed by replacing $\cc_X$ with $\Tilde{\cc}_X$; this is irrelevant to the result as the important outputs are the agents'.} Since routed CP maps form a compact closed category \cite{wilson2021composable}, $\mathcal{S}_{\Gamma}[(\cc_N)_N]$ follows the route $\mathcal{S}^\Rel_{\Gamma}[(\la_N^\aug)_N] = \La_\Gamma$. Since $\La_\Gamma$ is a function, this entails in particular that for any input state $\ketbra{\vec{i}}{\vec{i}}$, 

\be \mathcal{S}_{\Gamma}[(\cc_N)_N] \ketbra{\vec{i}}{\vec{i}} = \ketbra{\La_\Gamma(\vec{i})}{\La_\Gamma(\vec{i})} \, . \ee

We now suppose that the agents, who have access to classical settings $(I^q, Q_\send^q, Q^q_\rec)$ fix $i^q_\lost = Q_\send^q$  and $i^q_\cha = Q_\rec^q \oplus \bar{I}^q$, and that they pick $O^q := j^q_\pr$ as their guess for the sender's input.

Using the notations and result of Lemma \ref{lemma: p}, and supposing their inputs were initialised as per the games' rules, we then have an input $\vec{i} \in I_\pr^{Q_\send, Q_\rec \oplus \bar{I}^Q_\send}$, so $\La_\Ga(\vec{i})$'s value of $j^{Q_\rec}_\pr$ is $\delta^{Q_\rec}_{Q_\rec \oplus \bar{I}^Q_\send} = I^{Q_\send}$. We therefore have $O^{Q_\rec} = I^{Q_\send}$. This procedure defines a quantum instrument whose Kraus operators are given by (\ref{eq: instrument}).

\bibliographystyle{utphys}
\bibliography{refs}

\appendix

\end{document}